\documentclass[lettersize,journal,10pt,twocolumn]{IEEEtran}

\usepackage{booktabs}
\usepackage{wrapfig}
\usepackage{lipsum}
\usepackage{verbatim}
\usepackage{algorithm}
\usepackage{algorithmicx}
\usepackage{algpseudocode}
\usepackage{graphicx}
\usepackage{cite}
\usepackage[cmex10]{amsmath}
\usepackage{amsthm}
\usepackage{epstopdf}
\usepackage[font=small,labelfont=bf]{caption}
\usepackage{makecell}
\usepackage{arydshln}
\usepackage{multirow}
\usepackage{multicol}

\usepackage{hyperref}
\hypersetup{
    colorlinks=true,
    linkcolor=blue,
    filecolor=magenta,      
    urlcolor=cyan,
    pdftitle={Overleaf Example},
    pdfpagemode=FullScreen,
    }

\usepackage{nicefrac}
\usepackage{bm}
\def\0{{\mathbf 0}}
\def\1{{\mathbf 1}}

\def\b{{\mathbf b}}

\def\d{{\mathbf d}}

\def\g{{\mathbf g}}

\def\k{{\mathbf k}}

\def\m{{\mathbf m}}
\def\n{{\mathbf n}}
\def\o{{\mathbf o}}

\def\v{{\mathbf v}}  

\def\x{{\mathbf x}}

\def\A{{\mathbf A}}
\def\B{{\mathbf B}}
\def\C{{\mathbf C}}
\def\D{{\mathbf D}}

\def\H{{\mathbf H}}
\def\I{{\mathbf I}}

\def\M{{\mathbf M}}

\def\P{{\mathbf P}}
\def\Q{{\mathbf Q}}
\def\R{{\mathbf R}}

\def\X{{\mathbf X}}

\def\Z{{\mathbf Z}}

\def\ie{{\textit{i.e.}}}
\def\eg{{\textit{e.g.}}}

\def\cF{{\mathcal F}}
\def\cG{{\mathcal G}}

\def\cV{{\mathcal V}}

\def\bPhi{{\boldsymbol \Phi}}
\def\bPsi{{\boldsymbol \Psi}}
\def\bGamma{{\boldsymbol \Gamma}}

\def\bOmega{{\boldsymbol \Omega}} 
\def\bTheta{{\boldsymbol \Theta}}

\DeclareMathOperator*{\argmin}{arg\,min}
\usepackage{amsfonts}
\usepackage{float}

\usepackage{color}
\usepackage{xcolor}

\usepackage{mdwmath}
\usepackage{url}
\usepackage{amssymb}

\usepackage{caption}
\usepackage{subcaption}
\usepackage{float}

\usepackage{adjustbox} 
\usepackage{geometry}  

\newcommand{\red}[1] {\textcolor[rgb]{1.0,0.0,0.0}{{#1}}}

\begin{document}

%

\title{Deep Unrolling of Sparsity-Induced RDO for 3D Point Cloud Attribute Coding}
\author{
\IEEEauthorblockN{Tam Thuc Do, Philip A. Chou, \emph{Life Fellow, IEEE}, Gene Cheung, \emph{Fellow, IEEE}}
\renewcommand{\baselinestretch}{1.0}
\thanks{The work of G. Cheung was supported in part by the Natural Sciences and Engineering Research
Council of Canada (NSERC) RGPIN-2025-06252. \emph{(Corresponding author: Tam Thuc Do.)}}
\thanks{{T. Do and G. Cheung are with the department of EECS, York University, 4700 Keele Street, Toronto, M3J 1P3, Canada (e-mail:\{tamthuc, genec\}@yorku.ca).}}
}
\maketitle
%
\begin{abstract}
We study the problem of lossy attribute compression, given encoded 3D point cloud geometry available at the decoder, in a multi-resolution B-spline projection framework. 
A target continuous 3D attribute function is first projected onto a sequence of nested subspaces $\cF^{(p)}_{l_0} \subseteq \cdots \subseteq \cF^{(p)}_{L}$, where $\cF^{(p)}_{l}$ is a family of functions spanned by a B-spline basis function of order $p$ at a chosen scale and its integer shifts.
The projected low-pass coefficients $F_l^*$ are computed via variable-complexity unrolling of a rate-distortion (RD) optimization algorithm into a feed-forward network, where the rate term is the sparsity-promoting $\ell_1$-norm.
Thus, the projection operation is end-to-end differentiable.
For a chosen coarse-to-fine predictor, the coefficients are then adjusted to account for the prediction from a lower-resolution to a higher-resolution, which is also optimized in a data-driven manner.
\end{abstract}
%

%
\maketitle

\section{Introduction}
\label{sec:intro}

\textit{Point clouds} --- unordered lists of discrete points in 3D space --- are a popular 3D imaging representation, where each point comprises a 3D location as well as other attributes, such as color (\eg, red, green and blue). 
Point cloud {\em attribute coding} is the problem of coding the attributes assuming that the geometry (3D point locations) is already coded, transmitted, and available at the decoder.  The availability of the geometry at the decoder \textit{a priori} allows the attributes to be coded with the proper inner product.  In this paper, we improve upon the volumetric approach to 3D point cloud attribute coding studied in \cite{ChouKK:20}.  To be specific, \cite{ChouKK:20} showed that the \textit{Region-Adaptive Hierarchical Transform} (RAHT) in MPEG G-PCC \cite{QueirozC:16,SandriCKQ:19,SchwarzEtAl:18,graziosi2020overview,liu2020study} is a special case of multi-resolution analysis and synthesis using piecewise constant B-splines of order $p=1$ (\ie, degree 0), and also extended it to $p>1$.  It was a primarily analytical framework, in that the $p>1$ case required geometry-dependent eigendecomposition and inversion of large matrices with complexity $O(N^3)$ at both the encoder and decoder, where $N$ is the number of points in the point cloud.
In contrast, in our paper we circumvent eigendecomposition and mitigate matrix inversion by unrolling \textit{conjugate gradient descent} (CGD) \cite{saad2003iterative}, truncated to a finite number of steps.  This results in complexity $O(N)$ linear feed-forward networks at both encoder and decoder that arbitrarily closely approximate the orthogonal transforms in generalized RAHT and are amenable to end-to-end parameter optimization in a data-driven manner.
In so doing, for the first time, we offer an implementation of generalized RAHT that is not only practical but trainable.



We then leverage that result in the core of the paper, in which we bridge classical linear transform coding and nonlinear transform coding \cite{BalleCMSJAHT:21}, using a fully interpretable deep nonlinear encoding network that is equivalent to a \textit{rate-distortion} (RD) optimization procedure.
Specifically, we keep the near-orthogonal synthesis transform of our earlier-derived $O(N)$ decoder, but replace the encoder with an optimization procedure that finds the coefficients that, when quantized and decoded, minimize a RD Lagrangian.  As the RD Lagrangian we use a squared-error distortion term combined with a sparsity rate term, which we show is naturally minimized by {\em proximal gradient descent} (PGD) \cite{parikh13}.  By unrolling PGD to a fixed number of steps, we have a nonlinear feed-forward encoder network that is $O(N)$, trainable, RD-optimal, and amenable to various complexity-performance tradeoffs.  While RD criteria have been commonly used in end-to-end loss functions for learning codecs built from off-the-shelf neural architectures (\eg, \cite{BalleCMSJAHT:21}), by using the unrolled RD optimization algorithm as the encoder, we have a deep encoding network with only a handful of parameters, leaving most relevant parameters at the decoder.
As a result, using our generalized RAHT decoder, we have a \textit{fully interpretable} encoder and decoder with orders of magnitude fewer parameters to learn than other deep network approaches.

In addition, while \cite{ChouKK:20} did not address prediction, which was one of the main innovations in MPEG G-PCC over plain RAHT ($p=1$) \cite{LasserreF:19}, in our work we show how to integrate prediction into our feed-forward networks, and also how to train the predictor.
Experiments show a 6--11\% reduction in bit rate relative to prediction in MPEG G-PCC.

In summary, the major contributions of this paper are:
\begin{itemize}
    \item We offer the first $O(N)$ implementation of generalized RAHT, using algorithm unrolling to obtain linear feed-forward encoding and decoding networks that are both practical and trainable.
    \item We bridge classical and nonlinear transform coding, using a fully interpretable $O(N)$ deep nonlinear encoding network that is equivalent to an unrolled rate-distortion optimization procedure for choosing coefficients of the $O(N)$ linear decoding network.
    \item For a given coarse-to-fine predictor --- a key innovation in MPEG G-PCC --- we show how to adapt our encoding/decoding networks so that the predictor and our networks are co-trainable.
\end{itemize}

The outline of the paper is as follows.
Section\;\ref{sec:related} reviews related work.
Section\;\ref{sec:prelim} defines our terminology and presents our theoretical framework for orthonormal transform coding with generalized RAHT.
Section\;\ref{sec:formulate} shows how we approximate generalized RAHT by linear feed-forward networks at both the encoder and decoder, derived using unrolled conjugate gradient descent.
Section\;\ref{sec:unrolled2} shows how we pivot away from orthogonal transform coding by keeping the decoder but replacing the encoder with a rate-distortion optimized encoder (tailored to the decoder), using a nonlinear feed-forward network equivalent to an unrolled proximal gradient descent.
Section\;\ref{sec:prediction} shows how prediction can be incorporated seamlessly into our coding framework.
Section\;\ref{sec:results} presents experimental results, including RD performance results, complexity tradeoffs, and parameter comparisons.
Section\;\ref{sec:conclude} concludes.

\section{Related Work}
\label{sec:related}

\subsection{Point Cloud Compression}

\subsubsection{Graph signal processing (GSP)}

Early work on point cloud attribute compression was based on GSP, \eg, \cite{ZhangFL:14,Cohen2016gt,liu2022hybrid}, which required $O(N^3)$ eigendecomposition to compute geometry-dependent transforms.  Our work is inspired by and similar to the $O(N)$ vertex domain filtering in GSP, \eg, \cite{Pavez2021}.

\subsubsection{Region adaptive hierarchical transforms (RAHT)}

RAHT \cite{QueirozC:16, SandriCKQ:19} was adopted as the core transform in the MPEG geometry-based point cloud codec (G-PCC) \cite{SchwarzEtAl:18,graziosi2020overview,liu2020study}.  RAHT was generalized in \cite{ChouKK:20} and shown to be volumetric.  Our work builds on \cite{ChouKK:20} as described in Sec.~\ref{sec:intro}.
Rate-distortion optimized skip coding of RAHT coefficients is described in \cite{wang2025rdoskip}.



\subsubsection{Radiance fields}

Radiance fields are volumetric functions representing density and color, which may be fit to a set of existing views and rendered into novel views using volumetric rendering \cite{KajiyaV:84}.  Neural radiance fields (NeRFs) \cite{MildenhallSTBRN:20} use coordinate-based (or implicit) neural networks to represent these functions, while 3D Gaussian spatting (3DGS) uses an overcomplete set of Gaussians \cite{zwicker2002ewa, kerbl20233d}.  There has been some work on compressing these representations, \eg, \cite{niedermayr2024compressed, xie2024mesongs, chen2024hac, Xiangrui2025}.  Our work shows how B-splines can be used to represent and code such volumetric functions, though we focus on point clouds in this paper.



\subsubsection{Deep learning}

Neural architectures used for learned image compression, \eg, \cite{BalleCMSJAHT:21}, have been applied to point cloud geometry compression, \eg, \cite{guarda2020deep,tang2020deep}.  However, the same architectures have not been successfully applied to point cloud attribute compression, because they assume a signal over a dense grid. 
Nevertheless, there have been other black-box neural approaches to point cloud attribute compression, \eg, \cite{AlexiouTE:20, IsikCHJT:22, WangM:22, fang20223dac, shao2024diffusion, mao2025spac, huo2025sptrans, mao2025pcacgan, zhan2025}.  In contrast, we derive our nonlinear feed-forward neural network encoder from first principles as RD-optimized coefficient selection for a given decoder, and we derive our linear feed-forward neural network decoder directly from the assumption that the signal lies in a space spanned by B-splines of order $p$.

\subsubsection{Quality enhancement}

Both neural and non-neural methods have been used for quality enhancement, \eg, \cite{xing2023gqenet, guo2025pcegan, wei2025weiner}.  It is conceivable that these could be used to enhance our and other compression methods.



\subsection{Unrolling for Rate-Distortion Optimization}

\subsubsection{Rate-distortion optimization (RDO)}

The first codecs to be learned end-to-end using RDO were VQ-based \cite{ChouLG:89a}.  It was shown in \cite{ChouLG:89a} that an RD-optimal encoder is a parameter-free function (matched to the decoder and entropy code) that minimizes $D+\lambda R$ for each input.  Later, $D+\lambda R$ was used as the objective for neural codecs, \eg, \cite{BalleMSHJ:18,BalleCMSJAHT:21}.  However, neural codecs to date have used highly parameterized deep networks to approximate the minimization of the optimal encoder as a continuous function.  In contrast, while our encoder is also a deep network, it is the realization of an unrolled minimization procedure (determined by the parameters of the decoder and entropy code). This renders our codec fully interpretable and significantly reduces the number of parameters that need to be learned.

\subsubsection{Unrolling}

In some early works on algorithm unrolling \cite{monga21, gregor10} and more recently \cite{yu23nips, do2024interpretable}, neural networks are the result of various optimization problems, but so far none of them have been used for compression applications. To our knowledge, our work is the first to apply algorithm unrolling to compression.

\subsection{Improvements from Conference Versions}

The present paper represents the first archival-level publication of our work.  
However, it should be noted that some contributions have been previously published in earlier conference versions of this work.  
In particular, our theoretical foundation, unrolling of the matrix inverse, end-to-end learning, and prediction in the orthonormal case are treated to varying degrees in \cite{do2023volumetric, do2024volumetric, do2024learned}.  The present paper consolidates and unifies those contributions, and furthermore offers the following advances.

First, we cast rate-distortion optimization as $\ell_2$-norm minimization subject to an $\ell_1$-norm penalty (to promote sparse representation in the generalized RAHT transform domain), which is solvable using the iterative \textit{proximal gradient descent} (PGD) algorithm \cite{parikh13}. 
We interpret the RD-optimal encoder as an optimization procedure to find the optimal coefficients for a given decoder, sometimes called a decoder-only scheme.  
Besides offering significantly improved coding performance compared to \cite{do2023volumetric}, this approach does not require orthogonality and even allows overcompleteness of the decoder basis functions.  
Unrolling the iterative PGD algorithm into neural layers stacked back-to-back results in a learnable nonlinear feed-forward encoder, which offers variable encoding complexity, unlike the fixed computation in \cite{do2024learned, do2024volumetric}.  
We also use unrolled Conjugate Gradient Descent for matrix inversion in the decoder, which dramatically reduces computation compared to the unrolled Taylor expansion in \cite{do2023volumetric, do2024volumetric}.
Finally, we prepare the theoretical foundation for its extension to radiance field and Gaussian splat coding.













\section{Theoretical Foundation}
\label{sec:prelim}

In this section, we review and extend the theoretical framework for RAHT that we first elucidated in \cite{ChouKK:20, do2023volumetric, do2024volumetric, do2024learned}.  Our theoretical framework generalizes RAHT as used in MPEG G-PCC \cite{QueirozC:16,SandriCKQ:19,SchwarzEtAl:18,graziosi2020overview,liu2020study}, showing that it is a special case of multi-resolution orthogonal transform coding of volumetric functions, based on differentially encoding orthogonal projections of an original volumetric function onto a nested sequence of subspaces of volumetric B-splines of order $p$.  The section ends with the mathematical formulae for orthogonal transform coding using generalized RAHT of order $p$, denoted by RAHT($p$).

\subsection{Volumetric Functions and Coding Framework}

Here we introduce volumetric functions and our coding framework.  Volumetric functions are used to represent the variations of attributes across space.

Denote by $f \in \mathcal{F}:\mathbb{R}^3\to\mathbb{R}$
a \textit{volumetric function} in a linear vector space of functions $\cF$ that maps a 3D coordinate in $\mathbb{R}^3$ to an attribute in $\mathbb{R}$.
To
{\em code} or compress $f$ we fit it with a parametric function $f_V\in\cF_\cV\subseteq\cF$, where $\cV$ is a parameter space, quantize the parameters $V\in\cV$ as $\hat V$, and entropy code the quantized parameters $\hat V$.
We reproduce the original function $f$ as $f_{\hat V}$.
The distortion between $f$ and $f_{\hat V}$ will be measured as a squared norm,
\begin{align}
    D = \|f-f_{\hat V}\|^2 ,
    \label{eqn:D}
\end{align}
while the bit rate will be measured as an amount of information in $\hat V$,
\begin{align}
    R = rate(\hat V) = -\log_2 p(\hat V) ,
\end{align}
where $p(\hat V)$ models the probability of $\hat V$.
For a given Lagrange multiplier $\lambda>0$, we will select $V$ to minimize the Lagrangian
\begin{align}
    J = D + \lambda R .
\end{align}
The rate-distortion optimization aspect of our coding framework is new to this paper.

\subsection{Geometry-Dependent Norm}
\label{sec:geometry_dependent_norm}
Here we introduce the norm of a volumetric function, which is used to express the distortion \eqref{eqn:D} between original and reconstructed volumetric functions.   Different applications, such as point cloud compression and radiance field compression, require different distortion measures and hence different norms of the volumetric functions.  We define the norm flexibly in a way that admits different applications.  An important point will be that the norm used to measure the distortion of the attributes is generally geometry-dependent, unlike the norms used for coding images, video, and other signals with fixed domains.  Intuitively, this is because different parts of the attribute signal have different importance or accuracy requirements (\eg, based on perceptual sensitivity) depending on the underlying local geometry.

We express the norm in \eqref{eqn:D} as
\begin{align}
    \|f\| = \left(\sum_i (\xi_i^\top f)^2\right)^{1/2} ,
    \label{eqn:norm_of_f}
\end{align}
where $\xi_i^\top:\cF\to\mathbb{R}$
is a linear functional, \eg,
\begin{align}
    \xi_i^\top f = \int \xi_i(\x) f(\x) d\x ,
\end{align}
chosen to suit the application as well as the geometry of the signal.
Two applications are of particular interest.

The first application is coding of point cloud attributes.  For this application, we choose the linear functional
\begin{align}
    \xi_i^\top f = \int \delta(\x-\x_i)f(\x)d\x = f(\x_i)
    \label{eqn:xi_f_for_point_clouds}
\end{align}
to be the value of $f$ at the $i$-th point of the point cloud, $\x_i$.
With this choice of $\xi_i^\top f$, the distortion $D=||f-f_{\hat{V}}||^2$ is the squared error between $f$ and $f_{\hat{V}}$ evaluated at the points $\{\x_i\}$ of the point cloud.

The second application is coding of radiance fields \cite{MildenhallSTBRN:20}.  
For that application, one may choose the linear functional
\begin{align}
    \xi_i^\top f = \int_{t_n}^{t_f} T_i(t)\sigma(\x_i(t))f(\x_i(t))dt ,
\end{align}
where $\x_i(t)=\o_i+t\d_i$ is the ray (through the $i$-th pixel) from an observer at position $\o_i$ in direction $\d_i$, $t_n$ and $t_f$ are near and far bounds along the ray, $\sigma(\x)$ is the density at $\x$, $f(\x)$ is the attribute at $\x$, and
\begin{align}
    T_i(t) = \exp\left(-\int_{t_n}^t \sigma(\x_i(s)) ds \right)
\end{align}
is the accumulated transmittance along the ray from $t_n$ to $t$, \ie, the probability that the ray travels from $t_n$ to $t$ without hitting a particle \cite{KajiyaV:84,MildenhallSTBRN:20}.

In both applications, the support of the linear functionals $\xi_i^\top$ may be optionally enlarged to accommodate filtering and to avoid aliasing.

Also, in both applications the linear functionals $\xi_i^\top$ encapsulate the geometry of the signal to be coded.
In the point cloud application, the geometry is given by the point positions $\{\x_i\}$, while in the radiance field application, the geometry is given by the density function $\sigma(\x)$.
The
signal {\em geometry} is assumed to be known at the decoder prior to coding the {\em attributes}.
With the geometry fixed, $\xi_i^\top f$ is a linear functional of the attributes $f$.
Since the geometry generally changes with the signal to be coded, so does the norm \eqref{eqn:norm_of_f}.  This signal-dependent variability in the norm and the non-uniformity of the $\xi_i^\top$ are the main challenges in coding point clouds and radiance fields compared to coding images, video, and dense voxel grids, for which the geometry is regular and fixed.

We see from the above discussion that our coding framework accommodates both point clouds and radiance fields.  However, the present paper focuses on point clouds.
Application of the framework to radiance field compression is left to future work.

\subsection{Geometry-Dependent Inner Product}
\label{sec:inner_product}

Here we introduce the inner product $\langle f,g \rangle$ that induces the norm as $||f|| = \langle f,g \rangle^{-1/2}$.  This is important, as it puts additional structure on the function space $\cF$, elevating it from a normed vector space (a Banach space) to a vector space with inner product (a Hilbert space).  The inner product provides a notion of orthogonality, which is used to characterize the function in $\cF_\cV\subseteq\cF$ closest to $f\in\cF$ as the orthogonal projection of $f$ onto $\cF_\cV$.

Inducing the norm \eqref{eqn:norm_of_f} is the real-valued inner product
\begin{align}
    \langle f,g \rangle = \sum_i (\xi_i^\top f)(\xi_i^\top g)
    \label{eqn:inner_product}
\end{align}
for $f,g\in\cF$.
With $\xi_i^\top f$ defined as in \eqref{eqn:xi_f_for_point_clouds} for point clouds, this reduces to
\begin{align}
    \langle f,g \rangle = \sum_i f(\x_i)g(\x_i) .
    \label{eqn:inner_product_for_point_clouds}
\end{align}

Technically, to ensure that \eqref{eqn:inner_product} is an inner product and \eqref{eqn:norm_of_f} is a norm, each element of $\cF$ should be interpreted not a volumetric function $f$ but rather an {\em equivalence class} $[f]$ of volumetric functions \cite{Wiki_Equivalence_class}, where $f$ and $f'$ are in the same equivalence class $[f]$ if they agree on their linear functionals, \ie, $\xi_i^\top f = \xi_i^\top f'$ $\forall i$.  This interpretation ensures, for example, that an element of $\cF$ with zero norm is unique.  
It is easy to show that the space of equivalence classes remains a vector space, since $\alpha[f]=[\alpha f]$ and $[f]+[g]=[f+g]$ are well defined by the linearity of the functionals $\xi_i^\top$.  
However, for simplicity in the sequel we refer to the equivalence class $[f]$ through any of its members $f$.

With the inner product so defined, $\cF$ becomes a \textit{Hilbert space} \cite{Wiki_Hilbert_space}, and supports the usual affordances of Hilbert space such as orthogonal projection.

\subsection{Orthogonal Transform Coding}

Here we introduce orthogonal transform coding.
We consider the special case in which the reproduction function space $\cF_\cV$ is a linear vector space spanned by volumetric basis functions $\theta_1,\ldots,\theta_N$ that  are orthonormal, \ie, orthogonal with respect to the inner product \eqref{eqn:inner_product} and normalized with respect to the norm \eqref{eqn:norm_of_f}.
In other words, $\cF_\cV = \{\bTheta V:V\in\mathbb{R}^N\}$, where $V$ is a column-vector of coefficients, $\bTheta=[\theta_1,\ldots,\theta_N]$ is a row-vector of basis functions $\theta_n\in\cF$, and $\bTheta^\top\bTheta=\I$.
In the notation of this paper, $\bTheta^\top=[\langle\theta_n,\cdot\rangle]$ denotes the column-vector of functionals that are inner products with the basis functions, $\bTheta^\top f=[\langle\theta_n,f\rangle]$ denotes the column-vector of these functionals applied to $f$, and therefore $\bTheta^\top\bTheta$ denotes the $N\times N$ matrix of inner products of the basis functions with themselves.\footnote{
Our notation intentionally mimics standard matrix-vector notation, as if $f$, $\theta_1$, \ldots, $\theta_N$, and $f_V=\Theta V$ were finite-dimensional column vectors of length $M$, $\Theta=[\theta_1,\ldots,\theta_N]$ were an $M\times N$ matrix, and $\Theta^\top$ were its $N\times M$ transpose.  The casual reader is welcome to this intuitive interpretation.  However, the careful reader will note that we consistently treat $f$, $\theta_1$, \ldots, $\theta_N$, and $f_V$ as {\em volumetric functions}, $\Theta=[\theta_1,\ldots,\theta_N]$ as a {\em row-vector of functions}, and $\Theta^\top$ as a {\em column-vector of functionals}.  This allows our networks to represent and compute with volumetric functions such as $f$ indirectly through their inner products (such as the length-$N$ column vector $\Theta^\top f$) or their coefficients in a basis (such as the length-$N$ column vector of coefficients $V=(\Theta^\top\Theta)^{-1}\Theta^\top f$ of $f$ in the basis $\Theta$). This is similar to how computers process continuous-time waveforms in traditional digital signal processing.}

If $f_V=\bTheta V\in\cF_\cV$, then $V$ is the column vector of coefficients representing $f_V$ in the basis $\bTheta$.
If these coefficients are scalar quantized and entropy coded, then our coding framework reduces to {\em orthogonal transform coding} with a geometry-dependent inner product.



\subsection{Nested Subspaces of Volumetric Functions}

Here we extend orthogonal transform coding to {\em multi-resolution} orthogonal transform coding.

The function $f$ to be coded is orthogonally projected onto a nested sequence of \textit{function spaces} $\cF_{l_0}^{(p)},\cdots,\mathcal{F}_L^{(p)}$. 
Each $\cF_l^{(p)}$ is a linear vector space spanned by the volumetric B-spline functions $\phi_{l,\mathbf{n}}^{(p)}(\mathbf{x})=\phi_{0,\mathbf{0}}^{(p)}(2^l\mathbf{x}-\mathbf{n})$ of order $p$ and scale $l$ for offsets $\mathbf{n}\in\mathbb{Z}^3$.
That is, $\cF_l^{(p)}$ is the set of $C^{p-2}$-continuous volumetric functions that are tri-polynomial of degree $p-1$ over cubes $\{[2^{-l}([0,1)^3+\n):\n\in\mathbb{Z}^3\}$.
Since the cubes at level $l+1$ refine the cubes at level $l$, it is clear that any function in $\cF_l^{(p)}$ is also in $\cF_{l+1}^{(p)}$.
Thus, the function spaces are nested: $\cF_{l_0}^{(p)}\subseteq\cdots\subseteq\mathcal{F}_L^{(p)}$.

The orthogonal projection of $f$ onto each $\cF_l^{(p)}$ is the \textit{best approximation} $f_l^*$ of $f$ at level of detail $l$, \ie, $f_l^*=\arg\min_{f_l\in\cF_l^{(p)}}||f-f_l||^2$. 
The difference or {\em residual function} $g_l^* = f_{l+1}^* - f_l^*$ between consecutive approximations can be shown to lie in the linear function space $\mathcal{G}_l^{(p)}$ defined as the orthogonal complement of $\mathcal{F}_l^{(p)}$ in $\mathcal{F}_{l+1}^{(p)}$.
Then, dropping the superscript $p$ for brevity, one may write
\begin{align}
    \cF_L &= \cF_{l_0} \oplus \cG_{l_0} \oplus \cdots \oplus \cG_l \oplus \cdots \cG_{L-1}   \;\;\; \mbox{and}
    \label{eq:sum_of_subspaces} \\
    f_L^* &= f_{l_0}^* + g_{l_0}^* + \cdots + g_l^* + \cdots + g_{L-1}^* .
    \label{eqn:fg0_gL}
\end{align}
Thus, assuming $f_L^*(\mathbf{x}_i)=f(\mathbf{x}_i)$ for all $\mathbf{x}_i$, one may code $f$ by coding $f_{l_0}^*$, $g_{l_0+1}^*,\ldots,g_{L-1}^*$, which are each represented as coefficients in a designated basis to be quantized and entropy-coded. 
Moreover, $f$ may be transmitted and/or reconstructed progressively from coarse to fine via lower resolution representations $f_l^*$.

The following subsections will designate orthogonal bases for the nested sequence of function spaces $\cF_l$ and their complements $\cG_l$.

\subsection{Basis functions for $\mathcal{F}_l$ and their coefficients $F_l$}
\label{sec:lowpass_coeffs}

Here we provide a natural basis for each subspace $\cF_l$.
By definition $\cF_l$ is spanned by all B-spline functions $\phi_{l,\n}$ for $\n\in\mathbb{Z}^3$.
However, the dimensionality of $\cF_l$ (interpreted as the vector space of equivalence classes of linear combinations of $\phi_{l,\n}$ that agree on $\{\x_i\}$) is finite, say $\dim(\cF_l)=N_l\leq N$, where $N$ is the number of points in the point cloud.
This is because the number of B-spline functions $\phi_{l,\n}$ whose support contains a point of the point cloud is finite.  
Thus, from the spanning set of functions $\phi_{l,\n}$, one may select a finite subset of functions $\phi_{l,\n_k}$, $\n_k\in\mathcal{N}_l$, $|\mathcal{N}_l|=N_l$, as a basis for $\cF_l$.

Let $\bPhi_l=[\phi_{l,\mathbf{n}_k}]$ be the row-vector of $B$-spline basis functions for $\cF_l$, and let $\bPhi_l^\top=[\langle\phi_{l,\mathbf{n}_k},\cdot\rangle]$ be the column-vector of functionals that are inner products with the basis functions at scale $l$. 
Then, $\bPhi_l^\top f=[\langle\phi_{l,\mathbf{n}_k},f\rangle]$ is the column-vector of inner products of the basis functions with $f$, and $\bPhi_l^\top\bPhi_l$ is the matrix of inner products of the basis functions with themselves, namely the \textit{Gram matrix} for $\bPhi_l$ \cite{Wiki_Gram_matrix}.
The Gram matrix, which encapsulates the effects of point cloud geometry through the inner product, has full rank since $\bPhi_l$ is a basis.


Denote by $f_l^*$ the best approximation of the function $f$ in the function subspace $\mathcal{F}_l$, and let $f_l^*$ be represented by the column vector of {\em low-pass coefficients} $F_l^*$ in the basis $\bPhi_l$ such that
\begin{align}
    f_l^* = \bPhi_l F_l^* .
    \label{eqn:low_pass_synthesis}
\end{align}
Equation \eqref{eqn:low_pass_synthesis} can be considered a {\em synthesis operation}.
The low-pass coefficients $F_l^*$ can be calculated by solving the \textit{least squares} (LS) problem
\begin{align}
    F_l^* = \argmin_{F_l} \| f - \bPhi_l F_l\|_2^2
\end{align}
or equivalently solving the linear system
\begin{align}
    \tilde F_l^* \stackrel{\Delta}{=} \bPhi_l^\top f = ( \bPhi_l^\top \bPhi_l )F_l^*
    \label{eqn:lowpass_linear_system}
\end{align}
for
\begin{align}
    F_l^* = ( \bPhi_l^\top \bPhi_l )^{-1} \tilde F_l^* .
    \label{eqn:F_l_star}
\end{align}
The solution is unique (\ie, the inverse exists) since the Gram matrix has full rank.
Calculation of $\tilde F_l^*=\bPhi_l^\top f$ can be considered an {\em analysis operation}.

Now, since $\phi_{l,\n}\in\cF_l\subseteq\cF_{l+1}$, there exist coefficients $\{a_\k\}$ that express each B-spline function in $\cF_l$ in terms of the B-spline functions of $\cF_{l+1}$, namely
\begin{align}
    \phi_{l,\n}
    = \sum_{\k\in\mathbb{Z}^3} a_\k \phi_{l+1,2\n+\k}
    = \sum_{\k'\in\mathbb{Z}^3} a_{k'-2\n} \phi_{l+1,\k'}.
    \label{eqn:twoscale}
\end{align}
Indeed, these are convolutional coefficients, which depend neither on $l$ nor $n$, nor on the geometry.
Moreover, since the B-spline functions have bounded support, only a finite number of the coefficients $\{a_\k\}$ are non-zero.
In matrix form, 
\begin{align}
    \bPhi_l = \bPhi_{l+1} \A^\top_l ,
    \label{eqn:two_scale_equation_2}
\end{align}
where the row and columns of $\A^\top_l$ have been elided to retain only coefficients related to B-spline functions that have been selected as basis functions.
Thus, the analysis operation
\begin{align}
    \tilde{F}_l^* = \bPhi_l^\top f = \A_l \bPhi_{l+1}^\top f = \A_l \tilde{F}_{l+1}^*
    \label{eqn:tilde_F_l}
\end{align}
can be implemented by a sparse convolution of the coefficients from the next higher level of detail.
The sparse convolutions are parameterized by only a small number of coefficients $\{a_\k\}$, which are shared across levels.


At the highest level of detail, $L$, one may assume that the region of support of $\phi_{L, \n }$ is small enough that it contains only one point in the point cloud, and that its evaluation is also 1 at that point. 
This leads to $\bPhi_{L}^\top \bPhi_{L} = \I$, so that $\tilde F_L^* = \bPhi_{L}^\top f = F_L^* = [y_i]$, where $y_i = f(\x_i)$.

Then, for levels of detail from $l=L-1$ to $l=l_0$, $F_l^*$ can be computed from $\tilde F_l^*$ using \eqref{eqn:F_l_star}, where in turn $\tilde F_l^*$ can be computed from $\tilde F_{l+1}^*$ using \eqref{eqn:tilde_F_l}, and $\bPhi_l^\top\bPhi_l$ can be computed from $\bPhi_{l+1}^\top\bPhi_{l+1}$ using (from \eqref{eqn:two_scale_equation_2})
\begin{align}
    \bPhi_l^\top \bPhi_l = \A_l \bPhi_{l+1}^\top\bPhi_{l+1} \A^\top_l .
    \label{eqn:Phi_l_T_Phi_l}
\end{align}
Note that $\bPhi_l^\top \bPhi_l$ is a sparse matrix owing to the bounded support of the B-spline functions.
Hence, the non-zero coefficients of $\bPhi_l^\top \bPhi_l$ can be compacted into a dense matrix, say $[[\bPhi_l^\top \bPhi_l]]$, of size $N_l \times M$ with $M \ll N_l$.
Then, since \eqref{eqn:Phi_l_T_Phi_l} is linear in the elements of the matrix $\bPhi_{l+1}^\top\bPhi_{l+1}$, it can be computed using
\begin{align}
    [[\bPhi_l^\top \bPhi_l]] = \mathbf{\Gamma}_{\A_l} [[ \bPhi_{l+1}^\top\bPhi_{l+1} ]] ,
\end{align}
where the \textit{geometric attention} matrix $\mathbf{\Gamma}_{\A_l}$ is itself sparse,
its non-zero elements quadratic in the elements of $\A_l$ \cite{do2024volumetric}.

\subsection{Basis functions for $\mathcal{G}_l$ and their coefficients $G_l$}
\label{sec:highpass_coeffs}

Here we provide a natural basis for each subspace $\cG_l$.
By definition, $\mathcal{G}_l$ is the orthogonal complement of $\mathcal{F}_l$ in $\mathcal{F}_{l+1}$.
Hence, any basis for $\cG_l$ has $N_{l+1}-N_l$ elements.
Let $\bPsi_l$ be the row-vector of $N_{l+1}-N_l$ basis functions for $\mathcal{G}_l$, and express $\bPsi_l$ as a linear combination of $\bPhi_{l+1}$:
\begin{align}
    \bPsi_{l} = \bPhi_{l+1} \mathbf{Z}_l^\top . \label{eqn:psi_l_condition1}
\end{align}
Then choose $\Z_l^\top$ such that $\bPsi_l$ is orthogonal to $\bPhi_l$, \ie,
\begin{align}
    \bPhi_{l}^\top \bPsi_{l} = \mathbf{0}. \label{eqn:psi_l_condition2}
\end{align}
There are many choices for $\mathbf{Z}_l$ that satisfy \eqref{eqn:psi_l_condition2}, such as the following.
Let $\mathbf{I}_l^{b}$ be a {\em selection matrix} of size $({N}_{l+1} - {N}_l) \times {N}_{l+1}$ that selects a subset $b$ of basis vectors from level $l+1$,
\begin{equation}
    \mathbf{\Phi}_{l+1}^b = \mathbf{\Phi}_{l+1} \mathbf{I}_l^{b\top}.
\end{equation}
Then, $\mathbf{Z}_l^\top$ can be chosen as in \cite{do2024learned} to be
\begin{align}
    \mathbf{Z}_l^\top &= \left [ \mathbf{I} - \mathbf{A}_l^\top (\mathbf{\Phi}_{l}^\top\mathbf{\Phi}_{l})^{-1}\mathbf{A}_l
    (\mathbf{\Phi}_{l+1}^\top\mathbf{\Phi}_{l+1}) \right ] \mathbf{I}_l^{b\top} \\
    &= \mathbf{I}_l^{b\top} - \mathbf{A}_l^\top (\mathbf{\Phi}_{l}^\top\mathbf{\Phi}_{l})^{-1} \mathbf{\Phi}_{l}^\top\mathbf{\Phi}_{l+1}^b .
    \label{eq:Z_operation}
\end{align}
The intuition behind $\mathbf{Z}_l^\top$ can be understood from
\begin{align}
    \mathbf{\Psi}_{l} &= \mathbf{\Phi}_{l+1} \mathbf{Z}_l^\top \\
    &= \mathbf{\Phi}_{l+1}^b - \mathbf{\Phi}_l(\mathbf{\Phi}_{l}^\top\mathbf{\Phi}_{l})^{-1} \mathbf{\Phi}_{l}^\top\mathbf{\Phi}_{l+1}^b .
    \label{eqn:high_pass_basis_funcs}
\end{align}
As we can see, \eqref{eqn:high_pass_basis_funcs} first projects a selected set of basis functions $\mathbf{\Phi}_{l+1}^b$ orthogonally onto the lower-scale subspace $\cF_l$, and then uses the residuals as the basis functions for $\mathcal{G}_l$. 
It is straightforward to check that $\mathbf{\Psi}_{l}$ is orthogonal to $\mathbf{\Phi}_{l}$.

Now the coefficients $G_l^*$ of the residual function $g_l^*=f_{l+1}^*-f_l^*$ in the basis $\bPsi_l$ can be calculated by solving the linear system
\begin{align}
    \tilde G_l^* \stackrel{\Delta}{=} \bPsi_l^\top g_l^* = (\bPsi_l^\top\bPsi_l)G_l^*
\end{align}
for
\begin{align}
    G_l^* = (\bPsi_l^\top\bPsi_l)^{-1}\tilde G_l^* .
    \label{eqn:G_l_star}
\end{align}
where $\bPsi_l^\top\bPsi_l = \mathbf{Z}_l \mathbf{\Phi}_{l+1}^\top \mathbf{\Phi}_{l+1} \mathbf{Z}_l^\top $.

Analysis operation $\tilde G_l^*=\bPsi_l^\top g_l^*$ may be expressed as
\begin{align}
    \tilde G_l^*  
    & = \bPsi_l^\top(\bPhi_{l+1}F_{l+1}^*-\bPhi_lF_l^*) \\
    & = \Z_l\bPhi_{l+1}^\top\bPhi_{l+1}F_{l+1}^* \\
    & = \Z_l\tilde F_{l+1}^* .
    \label{eqn:tilde_G_l}
\end{align}

Then, for levels of detail from $l=L-1$ to $l=l_0$, $G_l^*$ can be computed by \eqref{eqn:G_l_star} with $\tilde G_l^*$ given by \eqref{eqn:tilde_G_l}.

\subsection{Reconstruction of $f$ from coefficients}

Here we show how to reconstruct $f$ from its coefficients in the bases defined above.

With coefficients $F_{l_0}^*,G_{l_0}^*,\ldots,G_{L-1}^*$
computed as described in the previous two subsections,
$f(\x_i)=[y_i]$ can be recovered as follows.

Since $f_{l+1}^* = f_l^* + g_l^*$ for each level of detail $l$, we have $\bPhi_{l+1}F_{l+1}^* = \bPhi_lF_l^* + \bPsi_lG_l^*$ or (applying the analysis operator $\bPhi_{l+1}^\top$)
$\bPhi_{l+1}^\top\bPhi_{l+1}F_{l+1}^* = \bPhi_{l+1}^\top\bPhi_{l+1}\A_l^\top F_l^* + \bPhi_{l+1}^\top\bPhi_{l+1}\Z_l^\top G_l^*$.
In other words,
\begin{align}
    F_{l+1}^* = \A_l^\top F_l^* + \Z_l^\top G_l^* .
    \label{eqn:synthesis_l}
\end{align}
Thus \eqref{eqn:synthesis_l} can be applied from level of detail $l=l_0$ to $l=L-1$ to recover $F_L^*=[y_i]$.

\subsection{Orthonormalized basis functions}

Here we show how to orthonormalize the basis functions.

While the bases $\bPhi_{l_0},\bPsi_{l_0},\ldots,\bPsi_{L-1}$ designated in subsections \ref{sec:lowpass_coeffs}--\ref{sec:highpass_coeffs} are orthogonal to each other, each basis is not orthonormal (\eg, the basis functions within $\bPsi_l$ are not orthogonal to each other).

However, the bases may be orthonormalized as done in \cite{do2023volumetric} by $\R_{\bPhi_l}=(\bPhi_l^\top \bPhi_l)^{-1/2}$ and $\R_{\bPsi_l}=(\bPsi_l^\top \bPsi_l)^{-1/2}$:
\begin{align}
    \overline{\bPhi}_{l} = \bPhi_l \R_{\bPhi_l} = \bPhi_l (\bPhi_l^\top \bPhi_l)^{-1/2} \\
    \overline{\bPsi}_{l} = \bPsi_l \R_{\bPsi_l} = \bPsi_l (\bPsi_l^\top \bPsi_l)^{-1/2} ,
\end{align}
such that $\overline{\bPhi}_l^\top\overline{\bPhi}_l=\I$ and $\overline{\bPsi}_l^\top\overline{\bPsi}_l=\I$.

Likewise the coefficients $F_l$ and $G_l$ may be orthonormalized:
\begin{align}
    \overline{F}_{l} = \R_{\bPhi_l}^{-1}F_l = (\bPhi_l^\top \bPhi_l)^{1/2} F_l
    \label{eqn:bar_F_l} \\
    \overline{G}_{l} = \R_{\bPsi_l}^{-1} G_l = (\bPsi_l^\top \bPsi_l)^{1/2} G_l ,
    \label{eqn:bar_G_l}
\end{align}
such that $\overline{\bPhi}_{l}\overline{F}_{l}=\bPhi_lF_l$ and $\overline{\bPsi}_{l}\overline{G}_{l}=\bPsi_lG_l$.

Now letting $\bTheta = [\overline{\bPhi}_{l_0}\;\overline{\bPsi}_{l_0}\;\cdots\;\overline{\bPsi}_{L-1}]$ be the row-vector of all orthonormalized basis functions at all levels, we have $\bTheta^\top\bTheta=\I$.  
Hence, the transform $V=\bTheta^\top f$ is orthonormal with inverse $f=\bTheta V$, where
\begin{align}
    V = \left[\begin{array}{c}
    \overline{F}_{l_0} \\
    \overline{G}_{l_0} \\
    \vdots \\
    \overline{G}_{L-1}
    \end{array}\right]
\end{align}
is the vector of orthonormalized coefficients.

\subsection{Encoder and decoder using orthonormal transform}

Finally, we summarize the actions of the encoder and decoder when using the orthonormal basis.  In particular, we show how to analyze an input function $f$ into the coefficients of the orthonormal basis, quantize the coefficients, and reconstruct the function as $\hat f$.

First, the encoder computes the coefficients $V$ from the values $y_i=f(\x_i)$, as follows.
Beginning with $\tilde F_L^* = [y_i]$, for levels of detail from $l=L-1$ to $l=l_0$, the encoder computes $\tilde F_l^*=\A_l\tilde F_{l+1}^*$ using \eqref{eqn:tilde_F_l}.

Then, combining \eqref{eqn:bar_F_l}, \eqref{eqn:F_l_star}, and \eqref{eqn:tilde_F_l}, the encoder computes $\overline{F}_{l_0}^* =  \R_{\bPhi_{l_0}}\A_{l_0}\tilde F_{l_0+1}^*$.

Finally, combining \eqref{eqn:bar_G_l}, \eqref{eqn:G_l_star}, and \eqref{eqn:tilde_G_l}, the encoder computes $\overline{G}_l^* =  \R_{\bPsi_l}\Z_l\tilde F_{l+1}^*$.

The coefficients $V$ are uniformly scalar quantized with a stepsize (say $\Delta$) matched to the desired bitrate, and entropy coded.
The decoder recovers the quantized coefficients as $\hat V$, which contains $\hat{\overline{F}}_{l_0}^*,\hat{\overline{G}}_{l_0}^*,\ldots,\hat{\overline{G}}_{L-1}^*$.

The decoder reconstructs the values $\hat f(\x_i)=\hat y_i$, where $\hat f=\bTheta\hat V$, as follows.
Beginning with $\hat F_{l_0}^*=\R_{\bPhi_{l_0}}\hat{\overline{F}}_{l_0}^*$ as in \eqref{eqn:bar_F_l},
for levels of detail from $l=l_0$ to $l=L-1$, with $\hat G_l^*=\R_{\bPsi_l}\hat{\overline{G}}_l^*$ as in \eqref{eqn:bar_G_l}, the decoder uses \eqref{eqn:synthesis_l} to compute
\begin{align}
    \hat F_{l+1}^* = \A_l^\top\hat F_l^* + \Z_l^\top\hat G_l^* .
\end{align}
Finally, $[\hat y_i]=\hat F_L^*$.

\begin{figure}
    \centering
    \includegraphics[width=\linewidth, trim={2.5cm 0.0cm 2.5cm 0.0cm},clip]{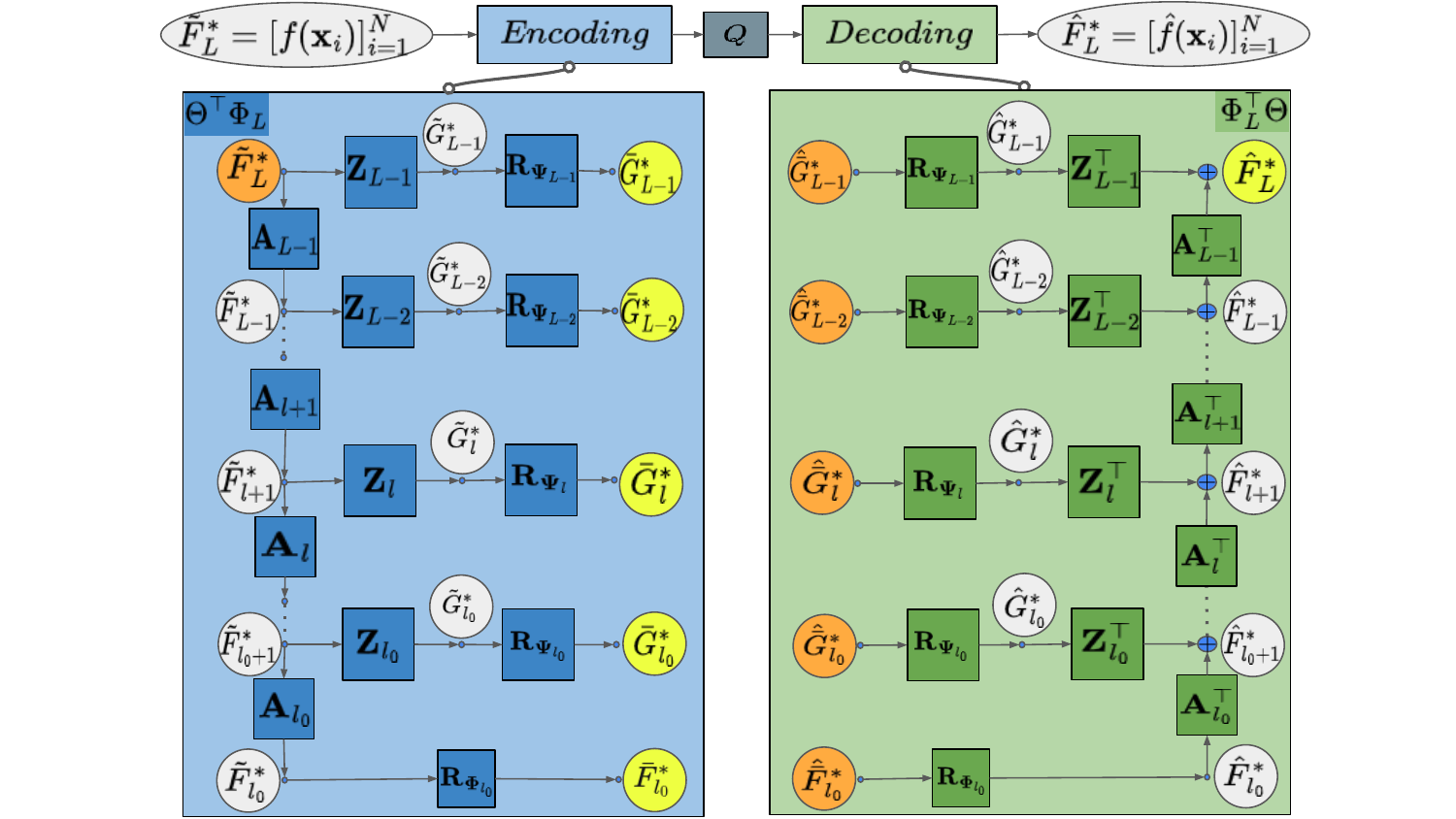}
    \caption{Ladder structure of encoder and decoder using orthonormal transform, analogous to a U-Net architecture, with quantization in between. The orange (and yellow) nodes correspond to the inputs (and outputs) of the blocks. Note that the encoding and decoding operations are expressed as $\bTheta^\top\bPhi_L$ and $\bPhi_L^\top \bTheta$, which are used in later diagrams as building blocks of our pipeline. Details of operation blocks $\mathbf{Z}_l$, $\mathbf{Z}_l^\top$, $\R_{\Psi_l}$ and $\R_{\Phi_{l_0}}$ are illustrated in later figures. The prediction module can also be included into the encoding and decoding operations, where the notations become $\bTheta_P^\top\bPhi_L$ and $\bPhi_L^\top \bTheta_P$.}
    \label{fig:encoder_decoder}
\end{figure}

The encoder and decoder can be viewed as a ladder structure illustrated in Fig.\;\ref{fig:encoder_decoder}. 

\section{Unrolled Generalized RAHT Transform}
\label{sec:formulate}

The previous section provided the mathematical formulae for orthogonal transform coding using RAHT($p$).  Unfortunately, the orthogonal analysis and synthesis transforms in RAHT($p$), illustrated in Fig.\;\ref{fig:encoder_decoder}, each have computational complexity $\Omega(N^2)$ in general, where $N$ is the number of points in the point cloud, which renders RAHT($p$) computationally infeasible for $p>1$.  In this section, we show how to approximate the orthogonal analysis and synthesis transforms with linear complexity, $O(N)$.

\begin{figure}
    \centering
    \includegraphics[width=\linewidth, trim={1.5cm 0.0cm 1.5cm 0.0cm},clip]{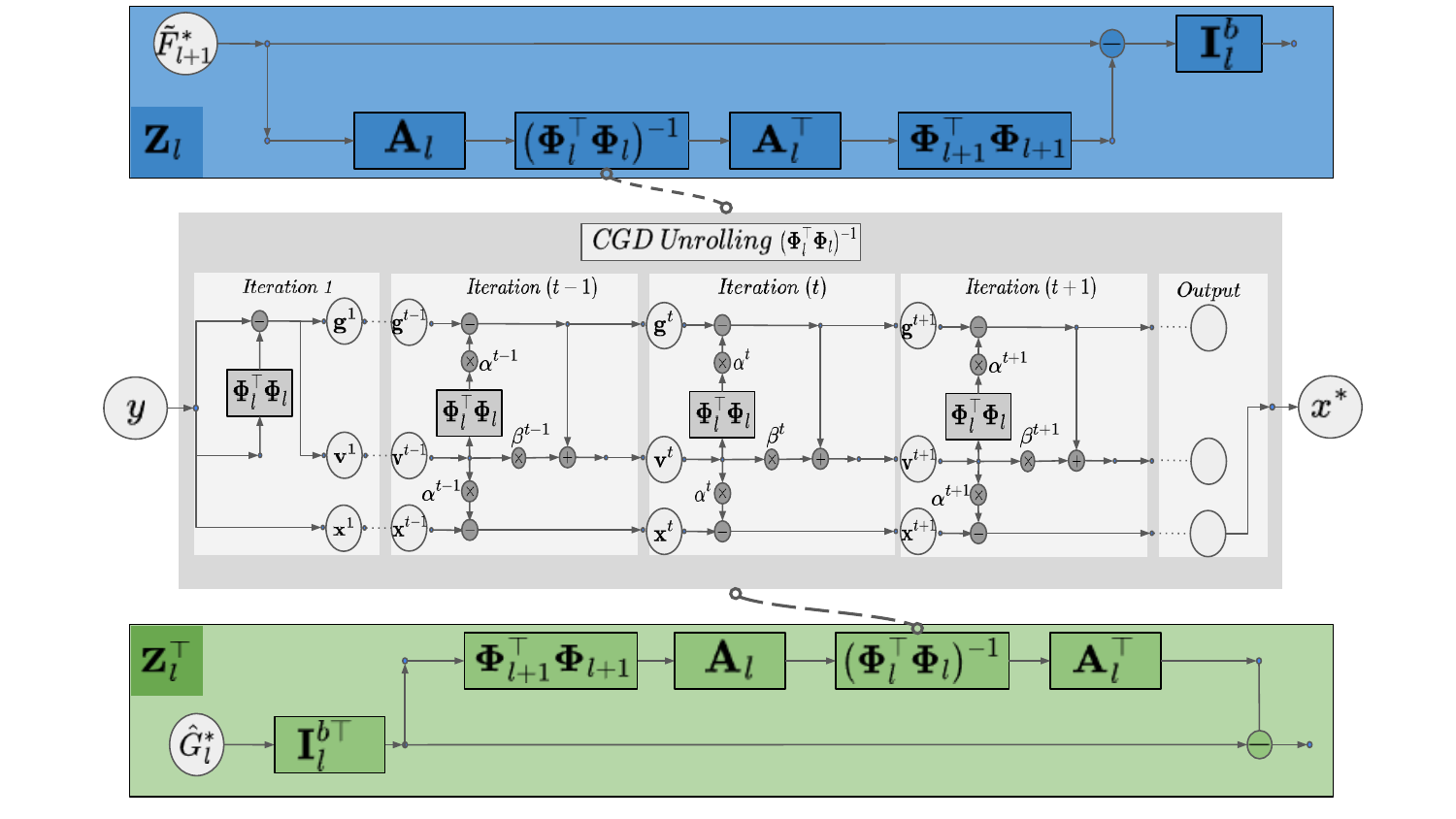}
    \caption{Sub-networks for $\mathbf{Z}_l$, $\mathbf{Z}_l^\top$, and $(\bPhi^\top\bPhi)^{-1}$ that are used in Fig.\;\ref{fig:encoder_decoder}. We describe Conjugate Gradient Descent Unrolling in Appendix~\ref{sec:unroll_cgd}. 
    } 
    \label{fig:Z_l_T}
\end{figure}

While the convolutional operator $\A_l$ in the analysis transform and the transpose convolutional operator $\A_l^\top$ in the synthesis transform have linear complexity $O(N)$, the other operators do not, in general.
For example, the operator $\Z_l^\top$, illustrated in Fig.~\ref{fig:Z_l_T}, is generally a dense matrix with complexity $\Omega(N^2)$, as it contains within it the operator $(\bPhi_l^\top\bPhi_l)^{-1}$, which in turn is dense as a matrix.
The same is true for the orthonormalization operators $\R_{\bPhi_l}=(\bPhi_l^\top\bPhi_l)^{-1/2}$ and $\R_{\bPsi_l}=(\bPsi_l^\top\bPsi_l)^{-1/2}=(\Z_l\bPhi_l^\top\bPhi_l\Z_l^\top)^{-1/2}$.

There is a notable exception to the general case when B-splines of order $p=1$ are used, as in the MPEG point cloud standard with RAHT($p=1$).  In that case, both $(\bPhi_l^\top\bPhi_l)$ and $(\bPsi_l^\top\bPsi_l)$ are diagonal, in which case their inverses and inverse square roots are also diagonal, and the operators are $O(N)$.
The situation is fundamentally due to the following well-known property of wavelets: the only orthogonal wavelets with finite support are the Haar wavelets \cite{Daubechies:92,VetterliK:95}. Nevertheless, in the general case, the operators in question can be approximated by unrolling them in a finite number of steps.

\subsection{Inverse operation}

Here, we show how to approximate the operator $(\bPhi_l^\top\bPhi_l)^{-1}$ with complexity $O(N)$, by unrolling conjugate gradient descent in a finite number of steps.

As Appendix~\ref{sec:taylor_inverse} shows, the Taylor expansion of the operator $\X^{-1}$ to $t+1$ terms is the degree-$t$ polynomial
\begin{align}
    \M^{(t)} = 2\mu \sum_{m=0}^t (\I-2\mu\X)^m
    \label{eqn:taylor_inverse} ,
\end{align} 
which converges to $\X^{-1}$ as $t\rightarrow\infty$ as long as $\mu<1/\lambda_{\max}(\X)$.
Interestingly, as Appendix~\ref{sec:taylor_inverse} also shows, \eqref{eqn:taylor_inverse} is the estimate for the matrix $\M$ that minimizes the squared matrix norm $||\X\M-\I||_{\P,\Q}^2$ after $t$ gradient descent steps with stepsize $\mu$ (where $\P$ and $\Q$ are chosen respectively as $\I$ and $\X^{-1}$).
This perspective leads us to improve the approximation even further using \textit{Conjugate Gradient Descent} (CGD), which is well-known for solving linear systems with positive-definite matrices and converging faster than gradient descent in such cases \cite{wiki_gcd}.
As Appendix~\ref{sec:unroll_cgd} shows,
the vector $F^{t}$ approximating $F=\X^{-1}\tilde F$ after $t>0$ CGD steps, starting with $F^{1}=\tilde F$ and $\g^1=\v^1=\tilde F - \X \tilde F$, is given by
\begin{align}
    F^{t}   &= F^{t-1} - \alpha_{t-1} \v^{t-1} \\
    \g^t &= \g^{t-1} - \alpha_{t-1} \X \v^{t-1} \\
    \v^{t} &= \g^t + \beta_{t-1} \v^{t-1} ,
\end{align}
where $\alpha_t$ and $\beta_t$ are analogous to step-size and momentum factors in the machine learning literature. Thus, we may approximate the operator $\X^{-1}$ by unrolling CGD to $t$ steps.
Since the matrix $\X=(\bPhi_l^\top\bPhi_l)$ is sparse, the operator $\X$ has complexity $O(N)$.
Thus, the operator $\X^{-1}$ approximated to $M_1$ steps has complexity $O(M_1N)$.
The operator may be implemented by a feed-forward network with $M_1$ layers as illustrated in Fig.~\ref{fig:Z_l_T}.



\subsection{Square-root inverse operation}

Here, we show how to approximate the operators $(\bPhi_l^\top\bPhi_l)^{-1/2}$ and $(\bPsi_l^\top\bPsi_l)^{-1/2}$ with complexity $O(N)$, by unrolling a Taylor series in a finite number of steps.

Appendix~\ref{sec:taylor_sqrt_inverse} shows that the Taylor expansion of the operator $\X^{-1/2}$ to $t+1$ terms is the degree-$t$ polynomial
\begin{align}
    \M^{(t)} = \sqrt{2\mu} \sum_{m=0}^t \frac{1\cdots(2m-1)}{2^m m!} (\I-2\mu\X)^m
    \label{eqn:taylor_sqrt_inverse} ,
\end{align}
which converges to $\X^{-1/2}$ as $t\rightarrow\infty$
as long as $\mu<1/\lambda_{\max}(\X)$.

\begin{figure}
    \centering
    \includegraphics[width=\linewidth, trim={0.5cm 3.8cm 0.5cm 3.8cm},clip]{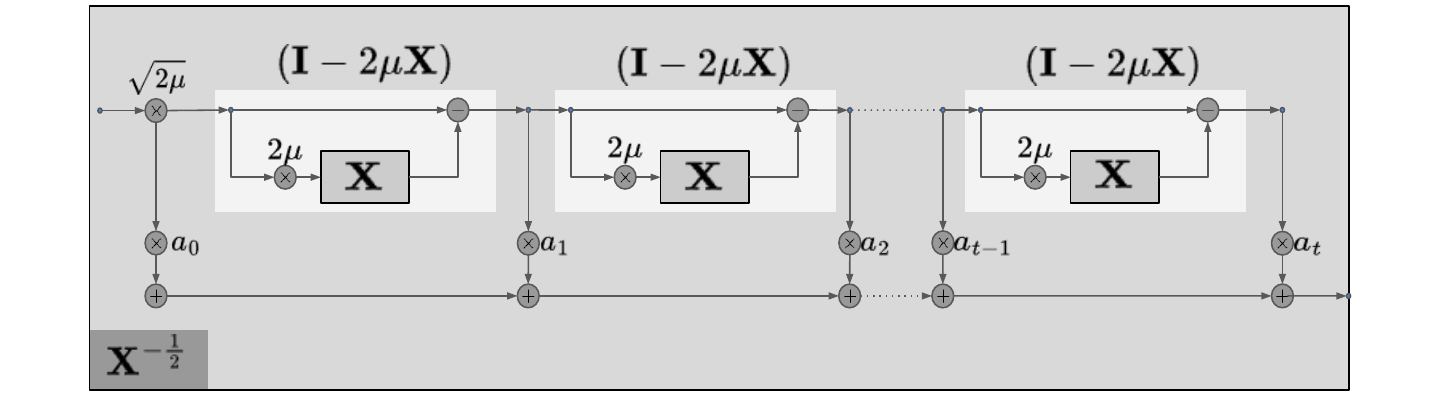}
    \caption{$M_2$-layer network approximating the ortho-normalization operator $\R_{\X} = \X^{-1/2}$, where the block $\X$ can be either the $O(N)$ operator $\bPhi_l^\top\bPhi_l$ or the $O(M_1N)$ subnetwork approximating the operator $\bPsi_l^\top\bPsi_l$. Hence, the operation blocks $\R_{\Psi_l}$ and $\R_{\Phi_{l_0}}$ in Fig.\;\ref{fig:encoder_decoder} are described.
    }
    \label{fig:taylor_sqrt_inverse}
\end{figure}

If $\X=(\bPhi_l^\top\bPhi_l)$, then as before the matrix $\X$ is sparse and the operator $\X$ has complexity $O(N)$.
Hence the operator $(\bPhi_l^\top\bPhi_l)^{-1/2}$ approximated to $M_2+1$ terms has complexity $O(M_2N)$.
The operator may be implemented by a feed-forward network with $M_2$ layers as illustrated in Fig.~\ref{fig:taylor_sqrt_inverse}.

However, if $\X=(\bPsi_l^\top\bPsi_l)$, then the matrix $\X$ is dense.
Indeed, Appendix~\ref{sec:Psi_Psi} shows in this case
\begin{align}
\mathbf{X}
& = \mathbf{I}_l^b\left[(\mathbf{\Phi}_{l+1}^T\mathbf{\Phi}_{l+1})\right. \\
& \left.-(\mathbf{\Phi}_{l+1}^T\mathbf{\Phi}_{l+1})\mathbf{A}_l^T(\mathbf{\Phi}_l^T\mathbf{\Phi}_l)^{-1}\mathbf{A}_l(\mathbf{\Phi}_{l+1}^T\mathbf{\Phi}_{l+1})\right]\mathbf{I}_l^{bT} ,
\nonumber
\end{align}
which itself involves the operator $(\bPhi_l^\top\bPhi_l)^{-1}$.
Nevertheless, $\X$ itself can be approximated by approximating the operator $(\bPhi_l^\top\bPhi_l)^{-1}$ with complexity $O(M_1N)$, as previously shown.
Therefore $\X^{-1/2}$ can be approximated by unrolling it into an outer loop with $M_2$ steps and an inner loop with $M_1$ steps, for an overall complexity of $O(M_1M_2N)$.
In particular, $(\bPsi_l^\top\bPsi_l)^{-1/2}$ can be implemented by a feed-forward network with $M_2$ layers of feedforward subnetworks each with $M_1$ layers.



\section{Unrolled Rate-Distortion Optimization}
\label{sec:unrolled2}

In the previous section, we showed how to implement the analysis and synthesis transforms, approximately, with complexity $O(N)$, by unrolling convergent series into finite numbers of steps.  The approximations result by truncating the series.  The approximations get ever more accurate as the length of the series grows, so there is a tradeoff between the closeness to orthonormality of the transforms and complexity.
In itself, approximate orthonormality does not necessarily lead to a significant degradation of the rate-distortion performance.  However, what does lead to a significant degradation is that the resulting synthesis transform no longer inverts the analysis transform, \ie, the transform no longer has the perfect reconstruction property.  In short, at some point, the PSNR saturates regardless of the bit rate.  The point of saturation begins at lower bit rates as the number of iterations $M_1$ and $M_2$ are reduced.

In this section, which is the core of our paper, we bridge from classical linear transform coding to nonlinear transform coding by proposing an alternative solution.
In our solution, orthonormality of the basis functions is not required; they may even be overcomplete.
Yet perfect reconstruction is possible 
as the encoder complexity (which is tunable) increases.

In our approach, we build on the $O(N)$ approximation introduced in the previous section.  Specifically, we re-use the linear, near-orthogonal decoder from the previous section with a fixed number of iterations $M_1$ and $M_2$ and decoder complexity $O(M_1M_2N)$.  Then, for the encoder, we use an optimization algorithm to choose the coefficient vector $V$ whose quantization $\hat V$, when decoded with the given decoder, minimizes our rate-distortion objective,
\begin{align}
    D+\lambda R = ||f-\bTheta\hat V||^2 + \lambda rate(\hat V) .
    \label{eqn:encoder_objective}
\end{align}
(Here, $\bTheta$ is the near-orthonormal basis of the given decoder.)
In the literature, using an optimization algorithm as the encoder is sometimes known as ``decoder-only'' compression, \eg, \cite{Park_2019_CVPR,mildenhall2020nerf,IsikCHJT:21,Dupont2021COINCW,Gordon:24}.

However, as ``decoder-only'' compression places no bound on the complexity of the optimization algorithm, we further propose to unroll the optimization algorithm into a finite number of steps.  Each of these steps will constitute a layer in a nonlinear feed-forward network of total complexity $O(N)$.

In this section, we show that minimizing the rate-distortion objective \eqref{eqn:encoder_objective} can be achieved efficiently using \textit{proximal gradient descent} (PGD) \cite{ParikhB:14}.  
Each iteration of the PGD simulates the decoder with complexity $O(M_1M_2N)$.  By unrolling the optimization algorithm into $M_3$ steps, the encoder becomes a nonlinear feed-forward network with $M_3$ layers of feed-forward subnetworks, each with $M_2$ layers of feed-forward subsubnetworks, each with $M_1$ layers.
Thus the unrolled encoder has complexity $O(M_1M_2M_3N)$.


Once both the encoder and decoder have been unrolled into feed-forward networks with a finite number of layers, it is possible to backpropagate through the layers in order to train all the parameters of the encoder and decoder.  Training will be discussed in Sec.~\ref{sec:results}.
Here note that the unrolled encoder has only a few parameters other than those in the decoder (which is simulated inside the encoder).  The few trainable parameters in the encoder are only those related to convergence of the optimization.

We now show how PGD emerges as a natural way to minimize the rate-distortion objective \eqref{eqn:encoder_objective}.

To begin, we need an expression for the bit rate, $rate(\hat V)$.
Recall that $\hat V$ is the vector of quantized coefficients containing $\hat{\overline{F}}_{l_0},\hat{\overline{G}}_{l_0},\ldots,\hat{\overline{G}}_{L-1}$.
(We continue to use the overbar notation even though the basis functions are no longer strictly orthonormal.)
Assume each coefficient, say $X$, is a zero-mean Laplacian random variable with mean absolute value $b=E[|X|]$, \ie, with density
\begin{align}
    p(x) = \frac{1}{2b}\exp\left(\frac{|x|}{b}\right)
\end{align}
and cumulative distibution function
\begin{align}
    C(t) & = \int_{-\infty}^t p(x)dx \\
    & = \left\{\begin{array}{ll}
    (1/2)\exp(t/b) & \mbox{if $t\leq 0$} \\
    1-(1/2)\exp(-t/b) & \mbox{if $t>0$}
    \end{array}\right. .
\end{align}
Then, when $X$ is uniformly scalar quantized with stepsize $\Delta$ as $\hat X=\Delta round(X/\Delta)$, we have
\begin{align}
    P\{\hat X=t\} = C(t+\Delta/2) - C(t-\Delta/2) ,
\end{align}
and hence the ideal number of bits used to encode the event $\{\hat X=t\}$ is \cite{CoverT:06}
\begin{align}
    \ell(t) = -\log P\{\hat X=t\} .
\end{align}
Appendix~\ref{sec:rate_term} shows that
\begin{align}
    \ell(t) \approx \frac{|t|}{b\ln2} - \log_2\frac{\Delta}{2b} .
\end{align}
Hence, assuming that all $N_{l_0}$ coefficients in $\hat{\overline{F}}_{l_0}$ have the same mean absolute value $a_{l_0}$, and that all $N_{l+1}-N_l$ coefficients in $\hat{\overline{G}}_l$ have the same mean absolute value $b_l$, we have
\begin{align}
    rate(\hat V) & = rate(\hat{\overline{F}}_{l_0}) + \sum_{l=l_0}^{L-1} rate(\hat{\overline{G}}_l) \\
    & = \frac{||\hat{\overline{F}}_{l_0}||_1}{a_{l_0}\ln2} + \sum_{l=l_0}^{L-1} \frac{||\hat{\overline{G}}_{l_0}||_1}{b_l\ln2} - K ,
\end{align}
where
\begin{align}
    K = N_{l_0}\log_2\frac{\Delta}{2a_{l_0}} + \sum_{l=l_0}^{L-1} (N_{l+1}-N_l)\log_2\frac{\Delta}{2b_l} .
\end{align}
Thus,
\begin{align}
    D+\lambda R = ||f-\bTheta\hat V||^2 + \lambda||\bGamma\hat V||_1 - \lambda K ,
    \label{eqn:encoder_objective2}
\end{align}
where $\bGamma= \mathrm{diag}([\gamma_i])$ is the diagonal matrix of elements $1/(a_{l_0}\ln2)$, $1/(b_{l_0}\ln2),\ldots,1/(b_{L-1}\ln2)$ respectively repeated $N_{l_0}$, $N_{l_0+1}-N_{l_0},\ldots,N_L-N_{L-1}$ times.

Since finding the integer vector $\hat V^*$ minimizing \eqref{eqn:encoder_objective2} would be a difficult integer problem, we relax the problem and instead find the real vector $V^*$ minimizing $||f-\bTheta V||^2+\lambda||\bGamma V||_1$, or
\begin{align}
    V^* = \argmin_{V=[v_i]} \left[||f-\bTheta V||^2 + \lambda\sum_i\gamma_i|v_i|\right] .
    \label{eqn:encoder_objective_relaxed}
\end{align}
Now,
$h(V) \stackrel{\Delta}{=} ||f-\bTheta V||^2$
is upper-bounded (majorized) at all $V$ by
\begin{align}
    h_\mu(V,V^{(t)}) \stackrel{\Delta}{=} & \; h(V^{(t)}) + \nabla h(V^{(t)})^\top(V-V^{(t)}) \nonumber \\
    & +\frac{1}{2\mu}||V-V^{(t)}||_2^2 ,
    \label{eqn:upper_bound}
\end{align}
for any $V^{(t)}$,
if $\mu$ is sufficiently small.
Hence, if $\mu$ is sufficiently small, \eqref{eqn:encoder_objective_relaxed} can be solved with the majorization-minimization algorithm
\begin{align}
    V^{(t+1)} & = \argmin_{V=[v_i]} \left[ h_\mu(V,V^{(t)}) + \lambda\sum_i\gamma_i|v_i|\right] .
    \label{eqn:majorization_minimization}
\end{align}
However, it can be seen from \eqref{eqn:upper_bound} that
\begin{align}
    & \mu h_\mu(V,V^{(t)}) \nonumber \\
    & = \mu\nabla h(V^{(t)})^{\top} V + \frac{1}{2}||V||^2 - V^{(t)\top} V + K' \\
    & = \frac{1}{2}||V - (V^{(t)}-\mu\nabla h(V^{(t)})||_2^2 + K'' ,
\end{align}
where $K'$ and $K''$ depend on $V^{(t)}$ but not on $V$.
Hence,
\begin{align}
    V^{(t+1)} = \argmin_{V=[v_i]} \left[\frac{1}{2}||V-U^{(t)}||^2 + \mu\lambda\sum_i\gamma_i|v_i|\right] ,
    \label{eqn:majorization_minimization2}
\end{align}
where
\begin{align}
    U^{(t)} & = V^{(t)}-\mu\nabla h(V^{(t)}) \\
    & = V^{(t)} + \mu\bTheta^\top(f-\bTheta V^{(t)}) .
    \label{eqn:gradient_step}
\end{align}
Note that $U^{(t)}$ can be interpreted as one gradient descent step away from $V^{(t)}$, with stepsize $\mu$.
Given $U^{(t)}$, since the argument in \eqref{eqn:majorization_minimization2} is now separable, $V^{(t+1)}$ is trivial to compute analytically and can be done in parallel.
Specifically, for each $i$,
\begin{align}
    v_i^{(t+1)}
    & = \argmin_{v} \left[\frac{1}{2\mu\lambda\gamma_i}(v-u_i^{(t)})^2 + |v|\right] \\
    & = \textit{Prox}_{\mu\lambda\gamma_i}(u_i^{(t)}) ,
    \label{eqn:scalar_proximal_operation}
\end{align}
where
\begin{align}
    \textit{Prox}_\tau(u) & \stackrel{\Delta}{=} \argmin_v \left[\frac{1}{2\tau}(v-u)^2+|v|\right] \\
    & = \left\{\begin{array}{ll}
    u - \tau & \mbox{if $u>\tau$} \\
    u + \tau & \mbox{if $u < -\tau$} \\
    0 & \mbox{if $|u|\leq\tau$}
    \end{array}\right.
    \label{eqn:shrinkage}
\end{align}
is a {\em proximal} operator, also known as a {\em shrinkage} operator in this case.
The shrinkage operator pulls its argument toward zero, thus encouraging sparse solutions.
Denote bt $V^{(t+1)} = \textit{Prox}_{\mu\lambda[\gamma_i]}(U^{(t)})$ the vector version of \eqref{eqn:scalar_proximal_operation}.
The equations for $U^{(t)}$ and $V^{(t+1)}$
are known as {\em proximal gradient descent} (PGD), as a proximal operator is applied to each gradient step.

\begin{figure}[t]
    \centering
    \includegraphics[width=\linewidth, trim={1.7cm 2.0cm 1.7cm 2.0cm},clip]{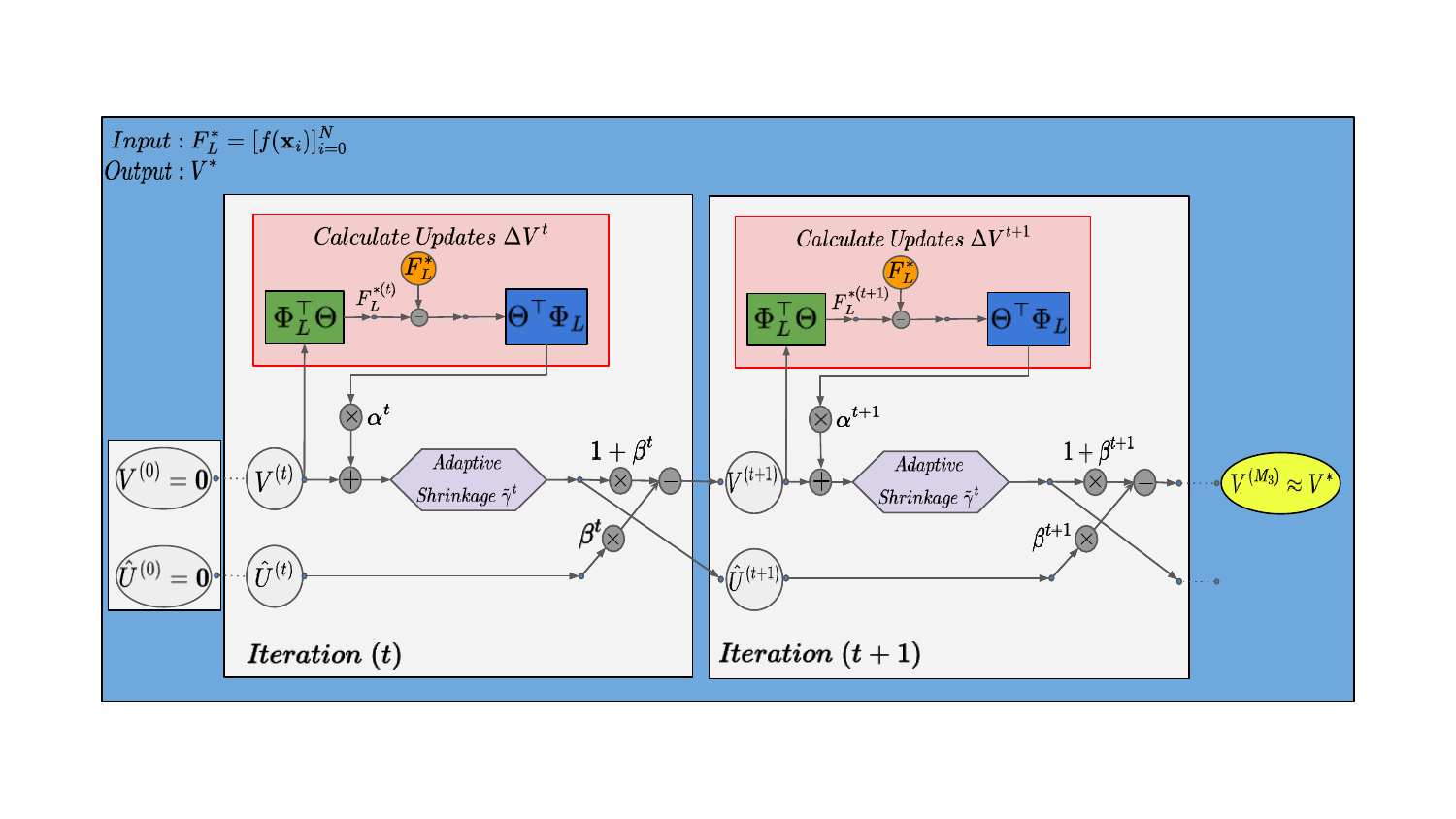}
    \caption{As described in eq.\eqref{eqn:update_step_v1}-\eqref{eq:delta_V_update}, the operations $\bTheta^\top\bPhi_L$ and $\bPhi_L^\top \bTheta$ described in Fig.\;\ref{fig:encoder_decoder} are now used to calculate the updates for the transformed coeficients then fed into the shinkage operation defined in eq.\;\eqref{eqn:shrinkage}. This whole block is our encoder as unrolled rate-distortion optimization, which is fully backpropagation-able and used in Fig.\;\ref{fig:learning_pipeline} as a crucial component in our learning pipeline.}
    \label{fig:feed-forward_encoder}
\end{figure}

We accelerate PGD using updating rules as in \cite{beck2009fast},
where a ``momentum" term is included in the iterations for faster convergence:
\begin{align}
     U^{(t + 1)} &= V^{(t)} + \alpha^t \bTheta^\top (f - \bTheta V^{(t)}) \label{eqn:update_step_v1} \\
     \hat{U}^{(t + 1)} &= \textit{Prox}_{\alpha^t\lambda[\gamma_i]}(U^{(t + 1)}) \\
     V^{(t + 1)} &= \hat{U}^{(t+1)} + \beta^t * \left ( \hat{U}^{(t+1)} - \hat{U}^{(t)} \right ) .
    \label{eqn:momentum_term}
\end{align}
Here, the parameters $\alpha^t$, $\beta^t$, $\gamma_i$ can be understood as learning rate, momentum, and $\ell_1$-regularizer coefficients in traditional optimization. In addition, by using the unrolling scheme, \eqref{eqn:update_step_v1}-\eqref{eqn:momentum_term} become consecutive blocks of a nonlinear neural network, as shown in Fig.~\ref{fig:feed-forward_encoder}.  The nonlinearity is the shrinkage operator \eqref{eqn:shrinkage}, which is a simple combination of rectified linear units (ReLUs).  In this way, these and other parameters become trainable.


It is worth noting that the update step $\Delta V^{(t)} = \bTheta^\top (f - \bTheta V^{(t)})$ in \eqref{eqn:update_step_v1} can be efficiently computed in the domain of samples $F_L^*$ of $f=\bPhi_L F_L^*$, as
\begin{align}
    \Delta V^{(t)}
    & = \bTheta^\top (\bPhi_L F_L^* - \bTheta V^{(t)}) \\
    & = \bTheta^\top \bPhi_L F_L^* - \bTheta^\top \bPhi_L \bPhi_L^\top \bTheta V^{(t)}) \\
    & = (\bTheta^\top \bPhi_L)[F_L^* - (\bPhi_L^\top \bTheta) V^{(t)}] ,
    \label{eq:delta_V_update}
\end{align}
as shown in Fig.~\ref{fig:feed-forward_encoder} with $F_L^{*(t)} = (\bPhi_L^\top \bTheta) V^{(t)}$.
Here we have used the identity $f=\bPhi_L \bPhi_L^\top f$ when $f\in\cF_L$.

\section{Prediction module}
\label{sec:prediction}

Prediction for RAHT($p=1$) coefficients was proposed in \cite{LasserreF:19}.  An improved causal context was proposed in \cite{PavezSQO:21}.  In our work (\cite{do2024learned, do2024volumetric} and this section), we extend to RAHT($p>1$), and use unrolled proximal gradient descent to find the rate-distortion optimal prediction residuals.

In the previous section, we showed that for a given linear decoder $\bTheta\hat V$ the encoder can find the $\hat V$ minimizing $D+\lambda R = ||f-\bTheta\hat V||^2+\lambda rate(\hat V)$ by minimizing $||f-\bTheta V||_2^2+\lambda||\bGamma V||_1$ using unrolled, accelerated PGD \eqref{eqn:update_step_v1}-\eqref{eqn:momentum_term}, and then quantizing $V$ to obtain $\hat V$.  Though we designed the decoder basis $\bTheta$ to be near-orthonormal, the PGD has no requirements on $\bTheta$.  Thus in this section, we are free to replace the decoder $\bTheta\hat V$ by a predictive decoder $\bTheta_P\hat V_P$ with basis $\bTheta_P=\bTheta\M$, where $V_P=\M^{-1}\hat V$ is a vector of prediction residuals, which are easier to code than $\hat V$.
To be specific, we let
\begin{align}
    V
    = \left[\begin{array}{c}
    \overline{F}_{l_0} \\
    \overline{G}_{l_0} \\
    \vdots \\
    \overline{G}_L
    \end{array}\right] ,
    \;\;\;
    V_P
    = \left[\begin{array}{c}
    \overline{F}_{l_0} \\
    \overline{G}_{l_0}'' \\
    \vdots \\
    \overline{G}_L''
    \end{array}\right] ,
\end{align}
where for each $l$, $\overline{G}_l=\overline{G}_l'+\overline{G}_l''$ are the normalized high-pass coefficients, $\overline{G}_l'=\B_l F_l$ are their predictions from the low-pass coefficients, and $\overline{G}_l''$ are the prediction residuals.

In our work, we choose
\begin{align}
    \B_l
    =
    \R_{\bPsi_{l}} \mathbf{Z}_{l} (\bPhi_{l+1}^\top\bPhi_{l+1})(\P^\top_{l} - \A^\top_{l}) .
\end{align}
This choice is motivated by the fact that if a matrix $\P_l^\top$ can predict $F_{l+1}$ perfectly from $F_l$, \ie, $\P^\top_{l}F_l = F_{l+1}$, then in the orthonormal case, $\overline{G}_l' = \overline{G}_l$ and $\overline{G}_l''=\0$ (and in the near-orthonormal case, $\overline{G}_l''\approx\0$).
On the other hand, if the prediction is weak, say $\P_l^\top = \A_l^\top$, then $\overline{G}_l$ is predicted to be $\overline{G}_l'=\0$, and $V_P$ reduces to $V$.  Other choices may also work.

Regardless of the exact choice of $\B_l$, we have
\begin{align}
    \left[\begin{array}{c}
    F_l \\
    \overline{G}_l
    \end{array}\right]
    =
    \left[\begin{array}{cc}
    \I & \0 \\
    \B_l & \I
    \end{array}\right]
    \left[\begin{array}{c}
    F_l \\
    \overline{G}_l''
    \end{array}\right] ,
    \label{eqn:Mp}
\end{align}
and its inverse
\begin{align}
    \left[\begin{array}{c}
    F_l \\
    \overline{G}_l''
    \end{array}\right]
    =
    \left[\begin{array}{cc}
    \I & \0 \\
    -\B_l & \I
    \end{array}\right]
    \left[\begin{array}{c}
    F_l \\
    \overline{G}_l
    \end{array}\right] .
    \label{eqn:Mp_inv}
\end{align}
It is shown in Appendix~\ref{sec:basis_changes} that $\M$ is a sequence of steps involving \eqref{eqn:Mp}, while $\M^{-1}$ involves \eqref{eqn:Mp_inv}.  Hence $\M$ and its inverse are straightforward to compute.

Now given $\M$, the encoder for the predictive decoder can find the $V_P$ minimizing $D+\lambda R = ||f-\bTheta_P\hat V_P||^2+\lambda rate(\hat V_P)$ by minimizing $||f-\bTheta_P V_P||_2^2+\lambda||\bGamma_P V_P||_1$ as usual using accelerated PGD:
\begin{align}
     U_P^{(t + 1)} &= V_P^{(t)} + \alpha^t \bTheta_P^\top (f - \bTheta_P V_P^{(t)}) \\
     \hat{U}_P^{(t + 1)} &= \textit{Prox}_{\alpha^t\lambda[\gamma_{Pi}]}(U_P^{(t + 1)}) \\
     V_P^{(t + 1)} &= \hat{U}_P^{(t+1)} + \beta^t * \left ( \hat{U}_P^{(t+1)} - \hat{U}_P^{(t)} \right ) ,
\end{align}
where $\bTheta_P=\bTheta\M$ and $\bTheta_P^\top=\M^\top\bTheta^\top$.
Note that while $\M^\top$ may be inconvenient to compute, it is shown in Appendix~\ref{sec:PGD_using_M_inverse} that $\M^{-1}$ may be used instead of $\M^\top$, possibly at the expense of slower convergence.

After the encoder unrolls accelerated PGD, it quantizes $V_P$ as $\hat V_P$ using a scalar quantizer with stepsize $\Delta$.  To quantize $V_P$, it is best practice for the encoder to limit error propagation with ``closed-loop'' quantization of $V_P$, \ie, with $\hat{\overline{G}}_l'' = Q(\overline{G}_l - \hat{\overline{G}}_l')$ (rather than $Q(\overline{G}_l - \overline{G}_l')$), so that $\hat{\overline{G}}_l'' = \overline{G}_l - \hat{\overline{G}}_l' + \epsilon_l$ and thus $\hat{\overline{G}}_l = \hat{\overline{G}}_l' + \hat{\overline{G}}_l'' = \overline{G}_l + \epsilon_l$, where $||\epsilon_l||_\infty<\Delta/2$.
However, we have not done so for the results in this paper.

The predictive decoder is simpler.  Once it receives $\hat V_P$, it obtains $\hat V=\M \hat V_P$ and reproduces $\hat f$ as $\bTheta\hat V$ as usual.

Experimental results in the next section show that prediction significantly improves rate-distortion performance.  However, the results also show that the gains are lower when the order of the base transform is higher.



\section{Experimental Results}
\label{sec:results}

\begin{figure}[t]
    \centering
    \includegraphics[width=\linewidth, trim={3cm 0.4cm 3cm 1.0cm},clip]{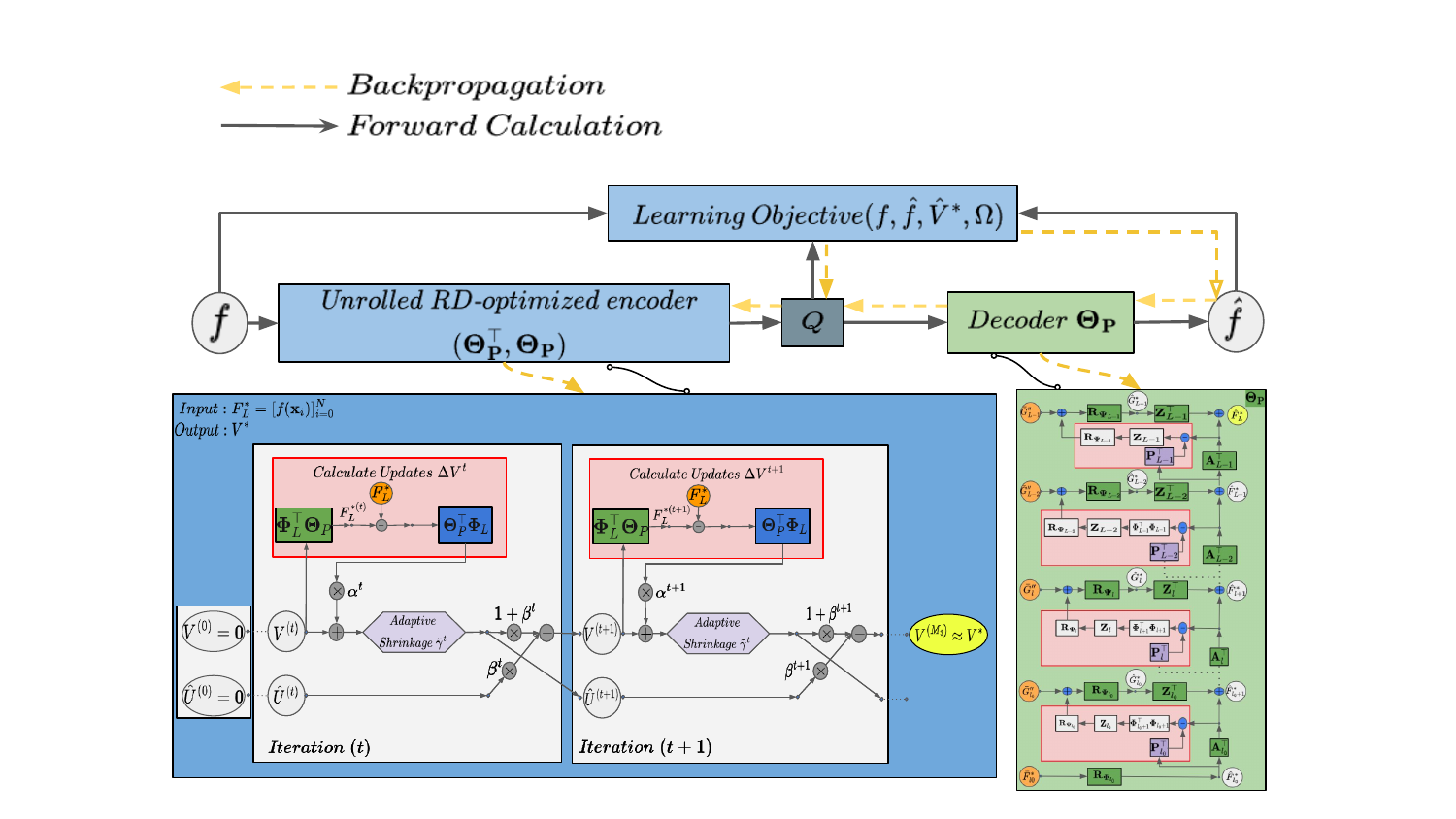}
    \caption{Our end-to-end training pipeline with parameters $\mathbf{\Omega}$. Different than Fig.\;\ref{fig:encoder_decoder}, our Encoder is an unrolled RD-optimization as described in Fig.\;\ref{fig:feed-forward_encoder}, where the operations  $\bTheta_P^\top\bPhi_L$ and $\bPhi_L^\top \bTheta_P$ are now included with a linear prediction module $P$ as described in Sec.\;\ref{sec:prediction} and Appendix~\ref{sec:PGD_using_M_inverse}.}
    \label{fig:learning_pipeline}
\end{figure}

\subsection{Unrolled neural network and learning objective}

In previous sections, we mathematically derived the neural architectures of sub-networks implementing various operations in our framework.  Figure \ref{fig:learning_pipeline} provides an overview of how the sub-networks are connected and trained end-to-end.  Table \ref{tab:trainale_parameter} summarizes the trainable parameters $\bOmega$. Our model has a total of about $742$ trainable parameters. The unrolled encoder, say $V_P=e(F_L;\bOmega)$, is a nonlinear function, differentiable almost everywhere, that produces coefficients $V_P=(\overline{F}_{l_0}, \overline{G}_{l_0}''$, ..., $\overline{G}_{L-1}'')$ from the point cloud samples $F_L$.  The decoder, say $\hat F_L=d(\hat V_P;\bOmega)$, is a linear function that reconstructs the point cloud samples $\hat F_L$ from the quantized coefficients.  The quantizer $\hat V_P=Q(V_P;\bOmega)$ is a uniform scalar quantizer with stepsize $\Delta$.  During training, it becomes a ``straight-through'' quantizer proxy, for which the gradients almost everywhere are unity rather than zero \cite{BalleCMSJAHT:21}.

During training, the loss function is the rate-distortion Lagrangian $J(\bOmega) = D(\bOmega) + \lambda R(\bOmega)$, where $D(\bOmega)$ is the mean squared reconstruction  error $D(\bOmega) = \frac{1}{N_L} \| F_L - \hat{F}_L \|_2^2$, $\lambda$ is a Lagrange multiplier, and $R(\bOmega)$ is the bit rate using an arithmetic coding proxy
\begin{align}
    R(V_P, \bOmega) &= -\frac{1}{N_L}\text{log}_2 p(\overline{F}_{l_0}; m_{F_{l_0}}, b_{F_{l_0}}, \Delta) 
    \nonumber \\
    &-\frac{1}{N_L}\sum_{l=l_0}^L \text{log}_2 p(\overline{G}''_{l_0}; m_l, b_l, \Delta) .
\end{align}
Here $p(\x;m,b,\Delta)=\prod p(x_i;m,b,\Delta)$, where
\begin{align}
    &p(x_i ; m, b, \Delta) 
    \nonumber \\
    &= \text{CDF}_{m, b}\left (x_i + \frac{\Delta}{2} \right) - \textit{CDF}_{m, b}\left (x_i - \frac{\Delta}{2} \right) ,
\end{align}
and $\textit{CDF}_{m, b}$ is the CDF of the Laplace distribution with location $m$ and mean absolute deviation $b$, and $\Delta$ is the quantization stepsize, all trainable parameters. 

\begin{table}[t]
\begin{center}
\begin{tabular}{||>{\raggedright}p{1.7cm}|>{\raggedright}p{2.0cm}|>{\raggedright}p{2.9cm}||} 
 \hline
 Parametrized operations & Parameters (per level) & Initialization (per level) \tabularnewline[0.1cm]
 \hline
 \hline&&\tabularnewline[-1em]
 $\A_l$, $\A^\top_l$ and $\bPhi_l^\top\bPhi_l$ &$(3\times 3\times 3)$-kernel with stride 2 in $\mathbb{Z}^3$ & RAHT(1) or RAHT(2) \tabularnewline[0.1cm]
 \hline&&\tabularnewline[-1em]
 Sub-network $(\bPhi_l^\top\bPhi_l)^{-1}$, \newline $\Z_l$, $\Z^\top_l$, and $\bPsi_l^\top\bPsi_l$ &CGD parameters \newline $[\alpha^{t}]^{M_1}_{t=1}$ and $[\beta^{t}]^{M_1}_{t=1}$ &$M_1$=5, $M_1$=15 for RAHT(1) and RAHT(2) repectively, with $\alpha^{t}$=0.99 and $\beta^{t}$=0.5 for all $t$\tabularnewline[0.1cm]
 \hline&&\tabularnewline[-1em]
 $\R_{\bPsi_l}$ and $\R_{\bPhi_l}$ &Taylor coefficients $[a_i]^{M_2}_{t=0}$ & Taylor expansion coefficients of $f(x)$=$\frac{1}{\sqrt{x}}$ \tabularnewline[0.1cm]
 \hline&&\tabularnewline[-1em]
 Prediction module\cite{do2024learned} $\mathbf{P}_l^\top$ &$(3\times 3\times 3)$-kernel &As described in \eqref{eq:predictor_init} \tabularnewline[0.1cm]
 \hline&&\tabularnewline[-1em]
 Unrolled Network RD-optimization $\ddagger$ &Update step size $[\alpha^{t}]^{M_3}_{t=1}$, momentum weight $[\beta^{t}]^{M_3}_{t=1}$ and shrinkage size $\tilde{\gamma}^{t}_{l}$ &$M_3$=5 (or $M_3$=10), with $\alpha^{t}$= 0.8 and $\beta^{t}$=0.1 for all $t$ and $\tilde{\gamma}^{t}_{l}=2e10^{-5}$ (or $\tilde{\gamma}^{t}_{l}=2e10^{-3}$) for RAHT(1) and RAHT(2) repectively\tabularnewline[0.1cm]
 \hline&&\tabularnewline[-1em]
 Learned Laplace distribution $\textit{CDF}_{m, b}$ &Location $m_l$, mean absolute deviation $b_l$, and quantization step size $\Delta$ &$m_l=0$ for $l> l_0$, $m_l=1$ for $l=l_0$, and $b_l=0.001$ for all $l$ \tabularnewline[0.1cm]
 \hline
\end{tabular}
\caption{Trainable Parameters: the right column in the table list additional parameters to define the sub-network (operations) on the left  column, e.g $(\bPhi_l^\top\bPhi_l)^{-1}$ required $\bPhi_l^\top\bPhi_l$ operations and additionally CGD parameters.
Totally, our model has about $\approx 750$ trainable parameters
}
\label{tab:trainale_parameter}
\end{center}
\end{table}

\begin{figure*}[t]
    \begin{subfigure}{0.237\textwidth}
        \centering
        \includegraphics[width=\textwidth,trim=0.0cm 0.0cm 0.0cm 0.0cm,clip]{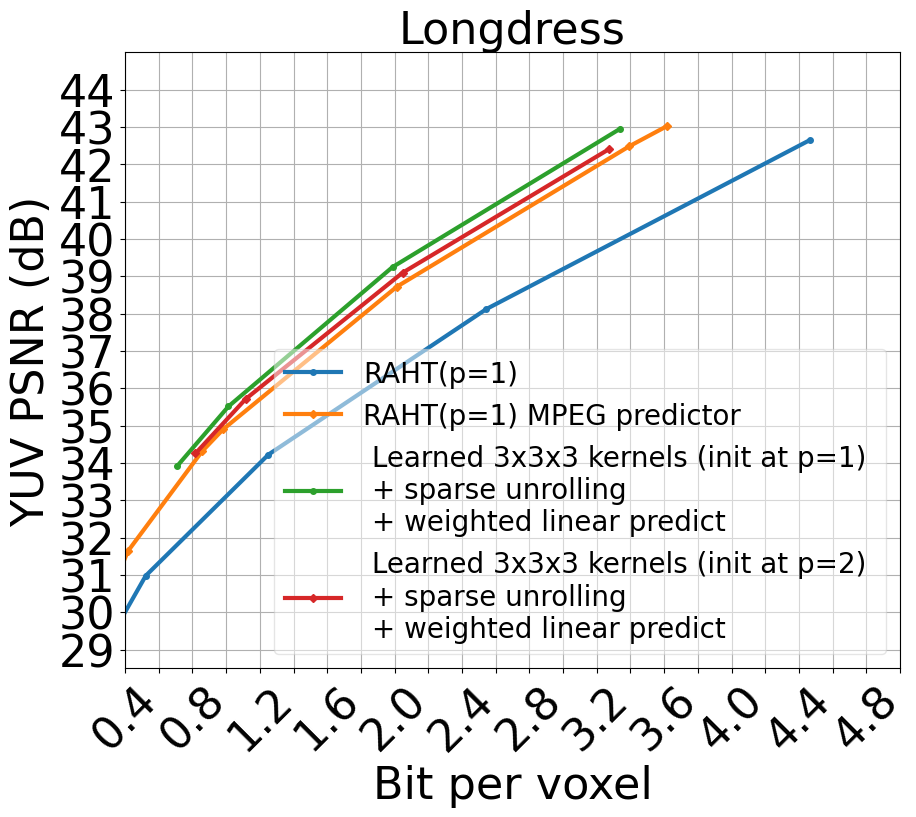}
        \caption{}
    \end{subfigure}
    \begin{subfigure}{0.25\textwidth}
        \centering
        \includegraphics[width=\textwidth,trim=0.0cm 0.0cm 0.0cm 0.0cm,clip]{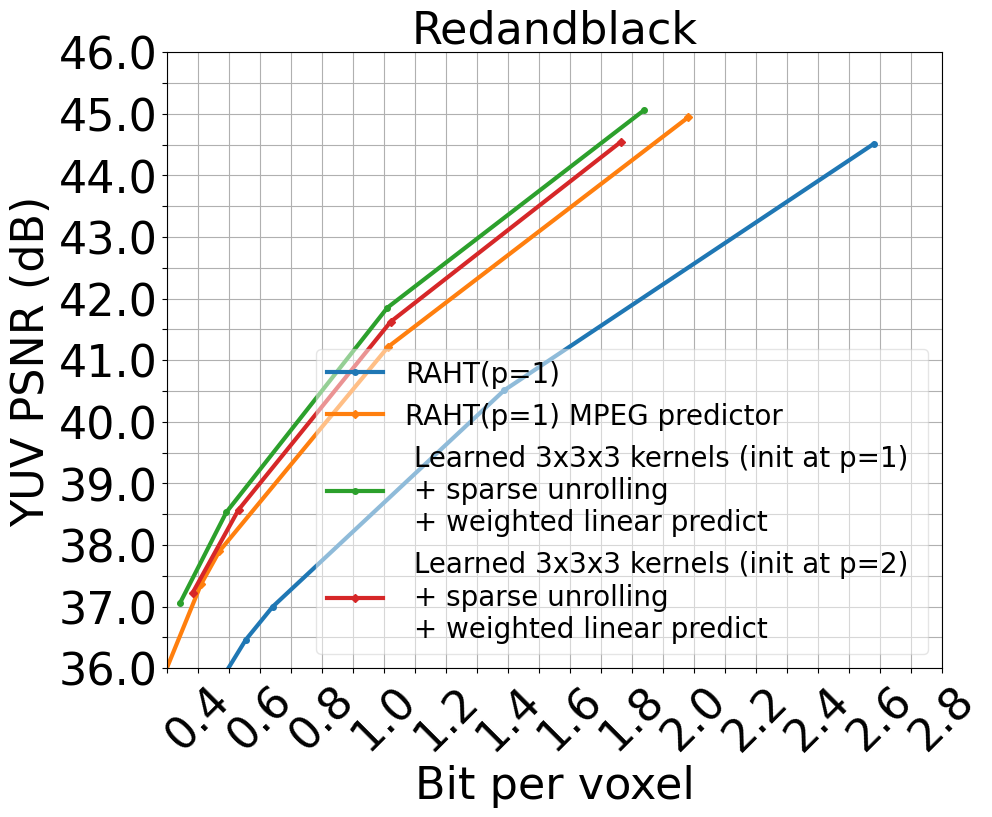}
        \caption{}
    \end{subfigure}
    \begin{subfigure}{0.25\textwidth}
        \centering
        \includegraphics[width=\textwidth,trim=0.0cm 0.0cm 0.0cm 0.0cm,clip]{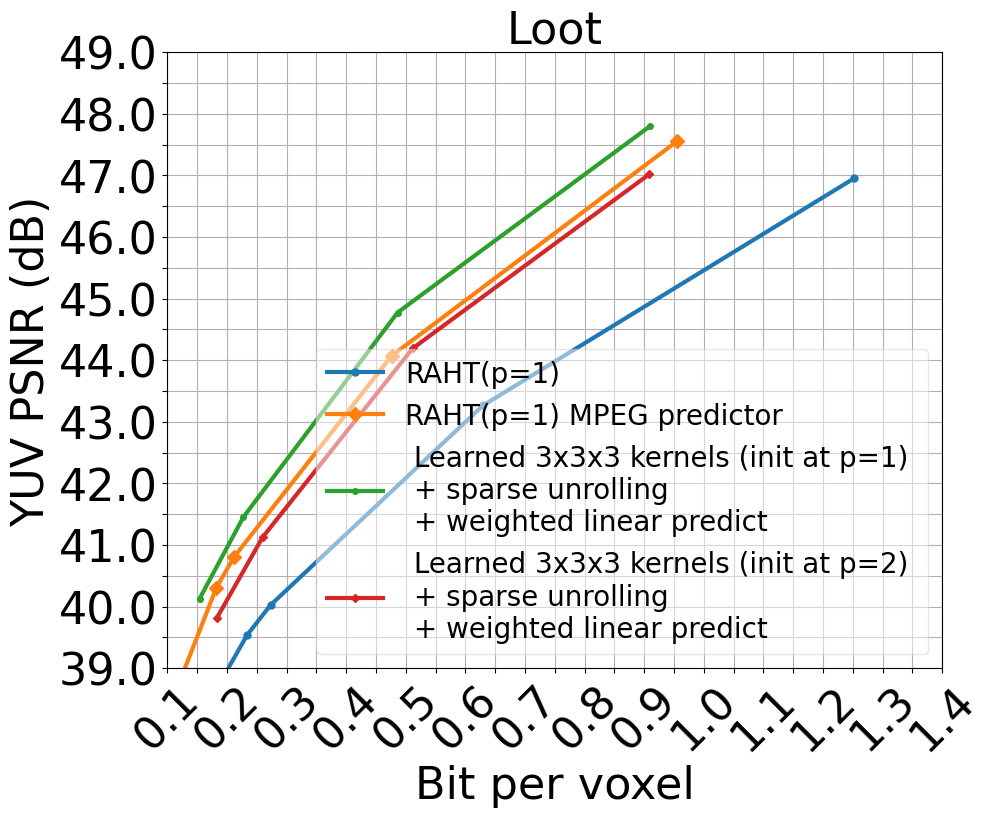}
        \caption{}
    \end{subfigure}
    \begin{subfigure}{0.247\textwidth}
        \centering
        \includegraphics[width=\textwidth,trim=0.0cm 0.0cm 0.0cm 0.0cm,clip]{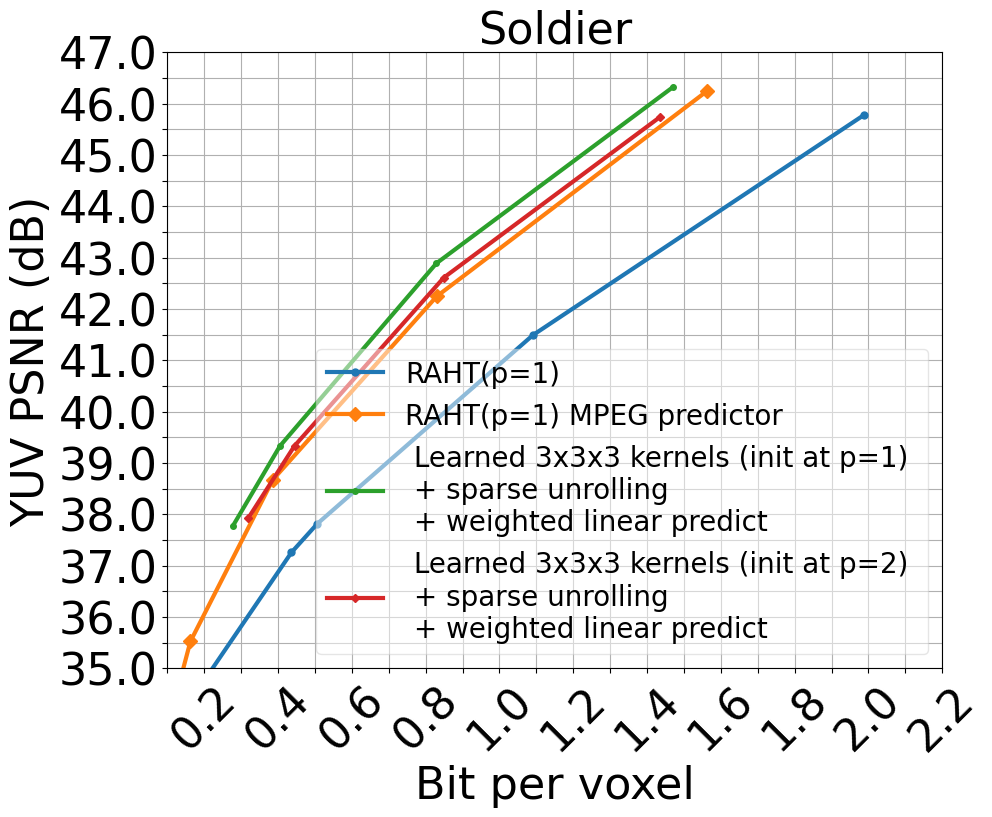}
        \caption{}
    \end{subfigure}
    \caption{Rate-Distortion curves: (a) \textit{Longdress}, (b) \textit{Redandblack}, (c) \textit{Loot}, (d) \textit{Soldier}}
    \label{fig:code_gain_performance_withP}
\end{figure*}

\begin{figure*}[t]
    \begin{subfigure}{0.237\textwidth}
        \centering
        \includegraphics[width=\textwidth,trim=0.0cm 0.0cm 0.0cm 0.0cm,clip]{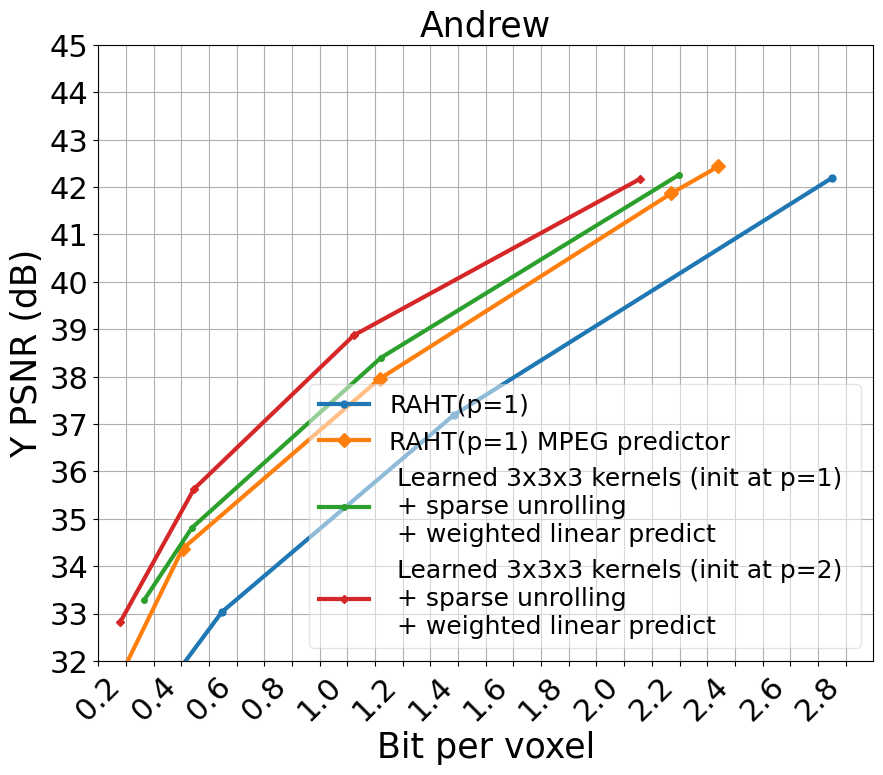}
        \caption{}
    \end{subfigure}
    \begin{subfigure}{0.25\textwidth}
        \centering
        \includegraphics[width=\textwidth,trim=0.0cm 0.0cm 0.0cm 0.0cm,clip]{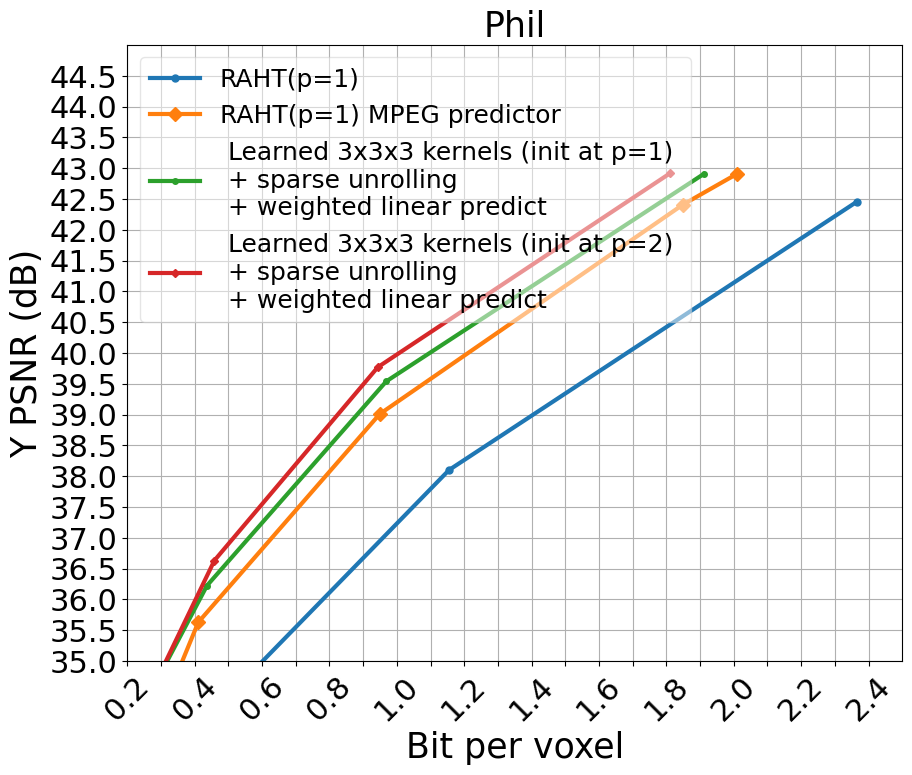}
        \caption{}
    \end{subfigure}
    \begin{subfigure}{0.25\textwidth}
        \centering
        \includegraphics[width=\textwidth,trim=0.0cm 0.0cm 0.0cm 0.0cm,clip]{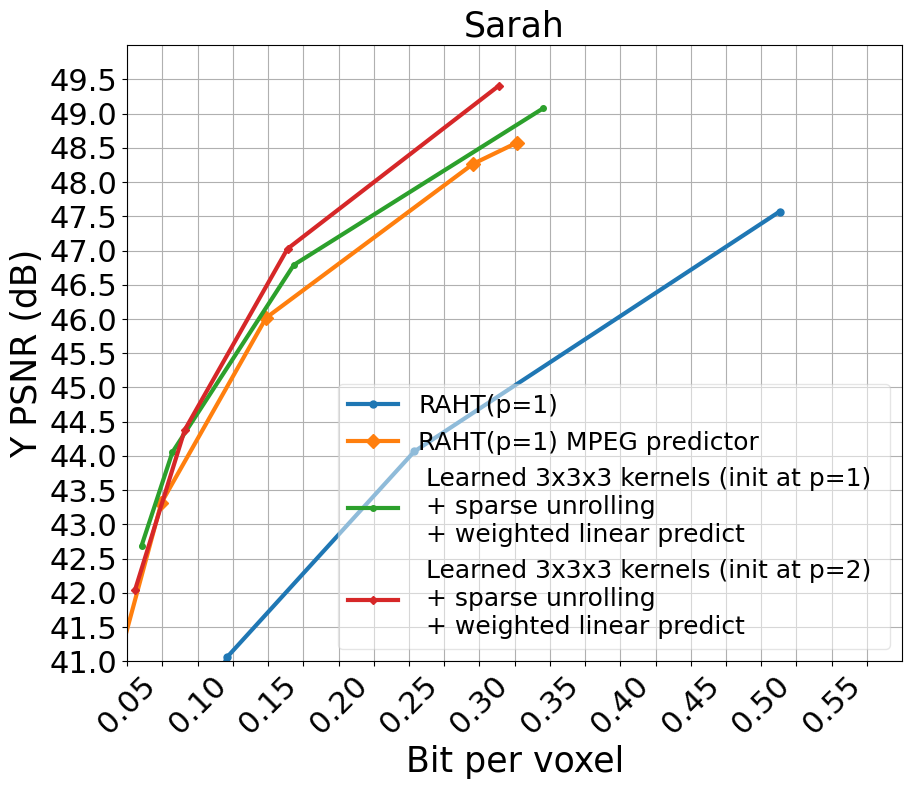}
        \caption{}
    \end{subfigure}
    \begin{subfigure}{0.247\textwidth}
        \centering
        \includegraphics[width=\textwidth,trim=0.0cm 0.0cm 0.0cm 0.0cm,clip]{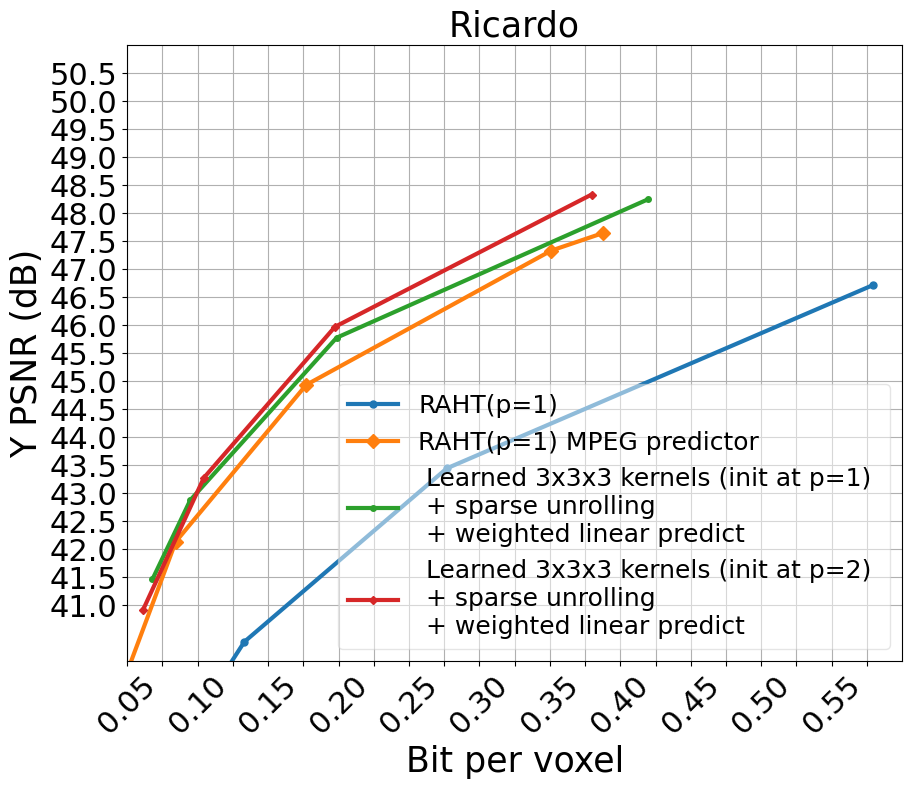}
        \caption{}
    \end{subfigure}
    \caption{Rate-Distortion curves with \textbf{Y-PSNR} for more dataset: (a) \textit{Andrew}, (b) \textit{Phil Chou}, (c) \textit{David}, (d) \textit{Ricardo}.}
    \label{fig:code_gain_performance_withP_more_data}
\end{figure*}

\subsection{Experiment Settings}

Our models were implemented in Python using Jax.  We used six levels of resolution from $l_0=4$ to $L=10$.  We trained fewer than $150$ parameters per level in various scenarios: a) without up-sampling predictor, b) with up-sampling predictor, and c) with various encoder/decoder complexity.  In all experiments, we initialized the $(3\times3\times3)$ kernels of the operations $\A_l$, $\A_l^\top$, and $\bPhi_l^\top\bPhi_l$ in two different ways, from the kernels for RAHT($p=1)$ and RAHT($p=2$), respectively.  The initialization of other parameters are briefly described in Table\;\ref{tab:trainale_parameter}.

Our dataset comprises more than 1000 point clouds voxelized to 10-bit resolution from 3D models (meshes) publicly available at sketchfab\footnote{https://sketchfab.com/blogs/community/sketchfab-launches-public-domain-dedication-for-3d-cultural-heritage/}. For training, we used only six levels of resolution created by $64\times64\times64$ crops centered on random point cloud points, resulting in about 14000 training examples. 
We used the Adam optimizer with learning rate 0.0001 in all configurations. 
For evaluation, we directly used the learned models on 10-bit resolution point clouds.

For evaluation, we used adaptive Run-Length Golomb-Rice (RLGR) entropy coding \cite{Malvar:2006} to calculate a realizable bit rate on the MPEG datasets {\em Longdress}, {\em Redandblack}, {\em Loot}, and {\em Soldier} \cite{dEonHMC:2017} and JPEG {\em Andrew}, {\em Phil}, {\em David}, and {\em Ricardo} \cite{JPEGpleno:2016}. 
For baselines, we estimated performance of the transform (with and without prediction) in the MPEG geometry-based point cloud codec (G-PCC) \cite{SchwarzEtAl:18, Wang2023predGPCC, GPCC:21} using RAHT($p=1$), which is the core transform of MPEG G-PCC, and the MPEG G-PCC predictor, where each coefficient $F^*_{l+1}[\mathbf{m_j}]$ is linearly predicted from a neighborhood of coefficients, $F^*_l[\mathbf{n}_i]$, $\mathbf{n}_i\in\mathcal{M}(\lceil\mathbf{m}_j/2\rceil) \subset \mathcal{N}_l$ as
\begin{eqnarray}
     \P_l^\top F_{l}^*=   \mathbf{D}^{-1} \mathbf{W} F_{l}^{*},
     \label{eq:predictor_init}
\end{eqnarray}
with $\mathbf{W}$ as a $N_{l+1} \times N_{l} $ matrix with entries  $w_{\mathbf{m}_j, \mathbf{n}_i}=1/d(2\mathbf{n}_i+\mathbf{1},\mathbf{m}_j+\mathbf{1}/2)$ when $(\mathbf{m}_j - 2\mathbf{n}_i) \in \{ 0, 1 \}^3 + \{ -1, 0, 1 \}^3$ and zero otherwise, $d$ is Euclidean distance, and $\D=diag([\sum_i w_{\m_j,\n_i}])$.

\subsection{Results Without Prediction}
\begin{figure}[h]
    \begin{subfigure}{0.23\textwidth}
        \centering
        \includegraphics[width=\textwidth,trim=13.0cm 5.0cm 15.0cm 5.0cm,clip]{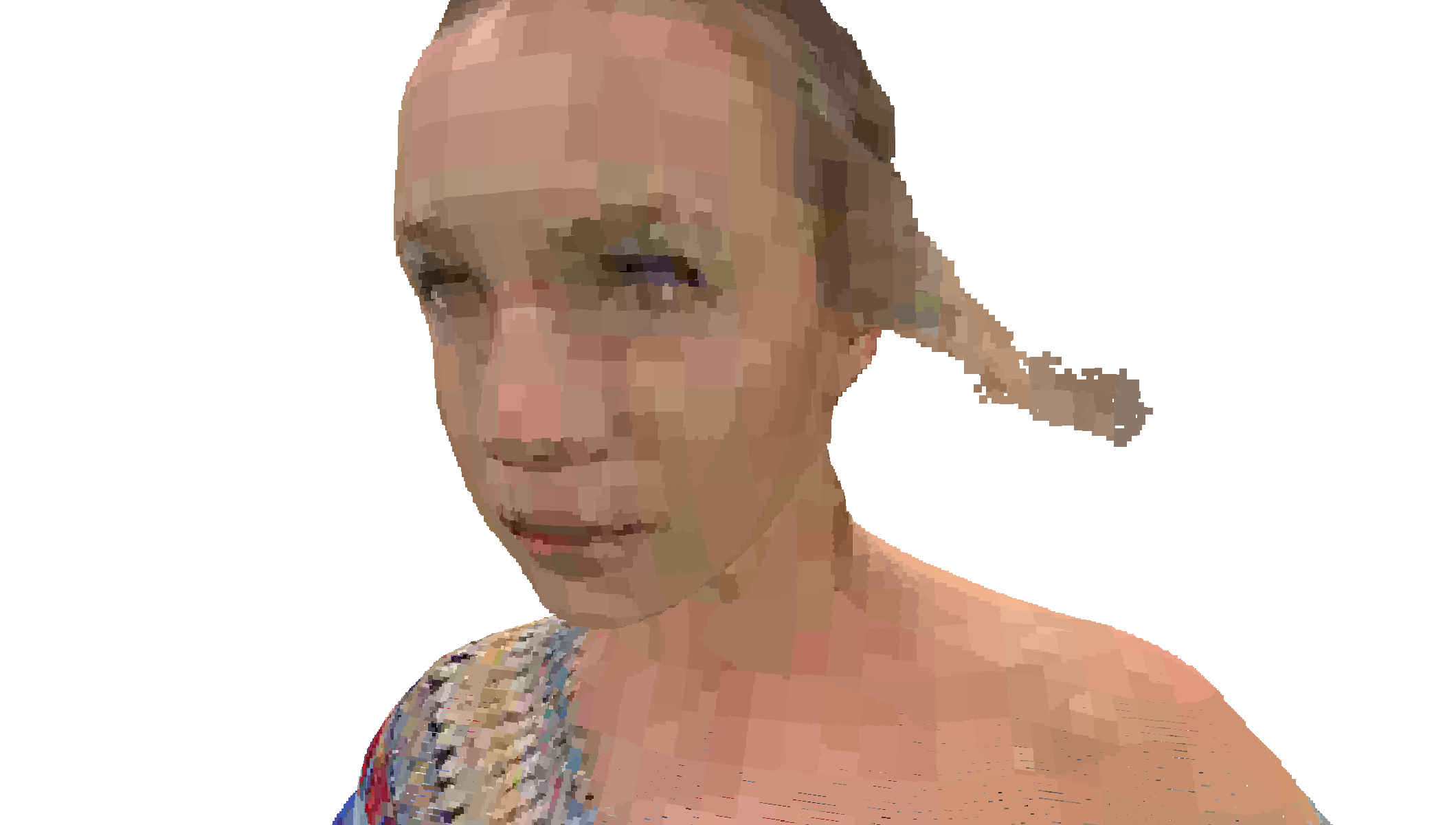}
    \end{subfigure}
    \begin{subfigure}{0.23\textwidth}
        \centering
        \includegraphics[width=\textwidth,trim=13.0cm 5.0cm 15.0cm 5.0cm,clip]{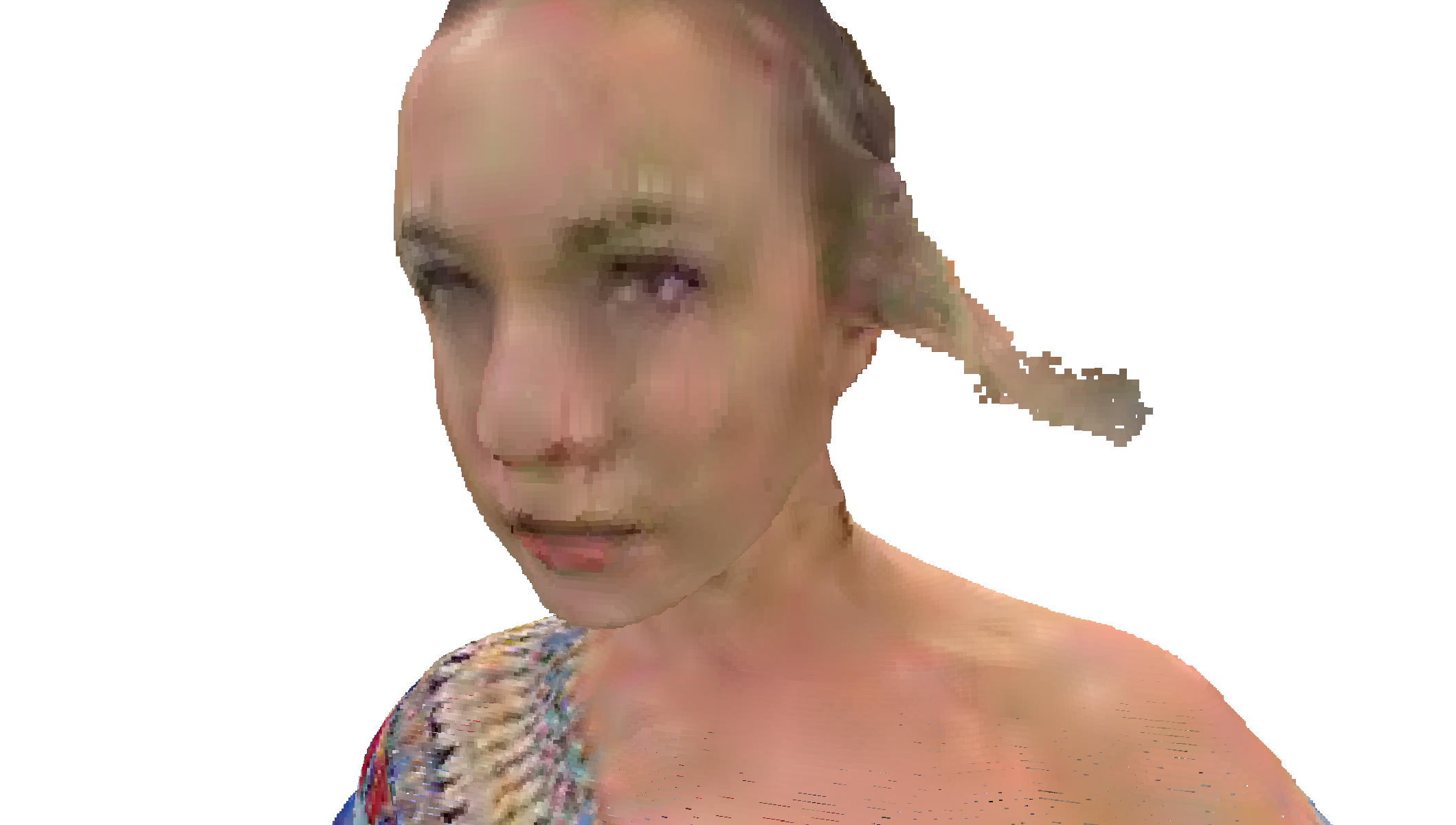}
    \end{subfigure}
    \caption{Reconstructed Point Cloud in case without prediction at $\sim$ 0.529 bits per point: (left) \textit{RAHT(p=1)} with 30.85dB, (right) \textit{RAHT(p=2)} with 31.84dB}
    \label{fig:recon_pc_wo_p}
\end{figure}

\begin{figure}[h]
    \begin{subfigure}{0.24\textwidth}
        \centering
        \includegraphics[width=\textwidth,trim=0.0cm 0.0cm 0.0cm 0.0cm,clip]{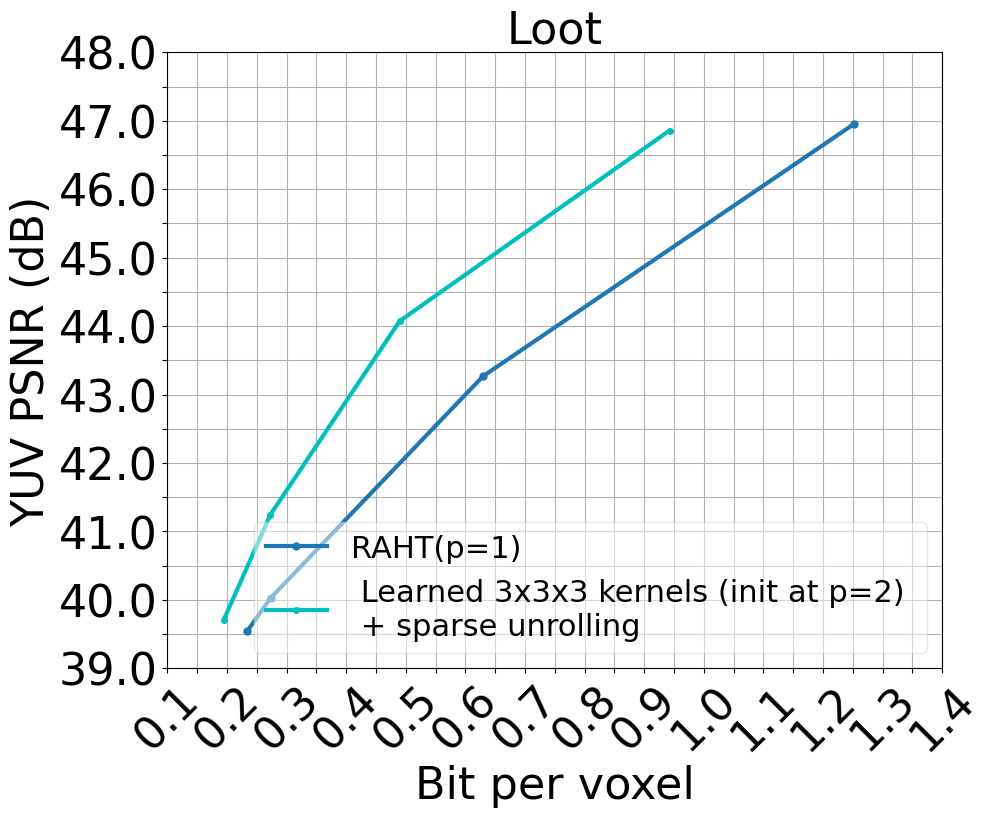}
    \end{subfigure}
    \begin{subfigure}{0.235\textwidth}
        \centering
        \includegraphics[width=\textwidth,trim=0.0cm 0.0cm 0.0cm 0.0cm,clip]{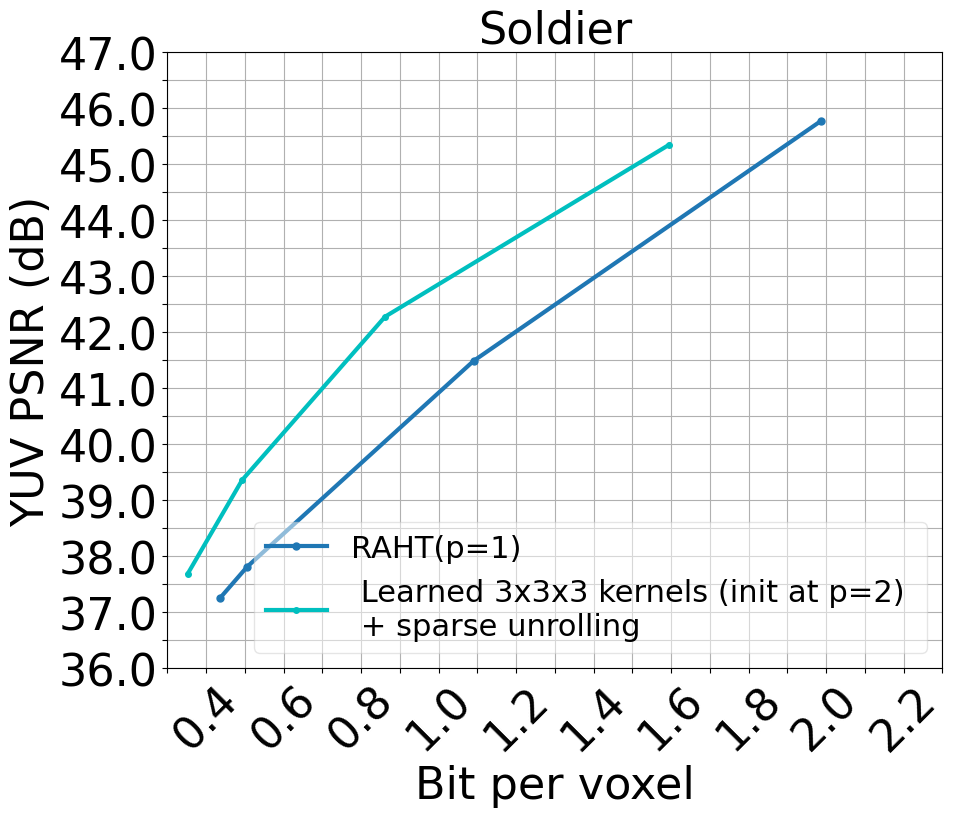}
    \end{subfigure}
    \caption{Rate-Distortion  curves: (left) \textit{Loot}, (right) \textit{Soldier}}
    \label{fig:code_gain_performance_wo_p}
    \vspace{-0.5cm}
\end{figure}

We first verify, without prediction, that higher-order B-splines ($p=2)$ can code point cloud attributes better than low-order B-splines ($p=1$). As shown in Fig.~\ref{fig:code_gain_performance_wo_p} and the reconstructed point clouds in Fig.~\ref{fig:recon_pc_wo_p}, higher-order B-splines lead to significantly better coding performance. The reduced bit rate is at least 20\% across all test points clouds. This is expected because functions in the subspace spanned by the higher-order B-splines are continuous and piecewise linear, guaranteeing a continuous transition of colors between blocks, whereas functions in the subspace spanned by low-order B-splines are only piecewise constant. 

\subsection{Results With Prediction}

Next, with up-sampling prediction available at each level of detail, we focus on the ability of our framework to learn. We tested our models by training with different initializations using up-sampling prediction. Figure~\ref{fig:code_gain_performance_withP} shows that our models exhibit significant coding gains over RAHT($p=1$) with up-sampling prediction: $6\%-11\%$ reduction in bit rate or $\sim 1$dB gain in YUV PSNR. Our coding improvements are largely attributable to the freedom to adjust the predictor weights at different levels and also the ability to learn the kernels for $\A_l$ that define the sequence of subspaces. 

We also observed that the performance of the learned models initialized at $p=2$ performs best with distortion metric Y-PSNR for datasets \textit{Andrew}, \textit{Phil}, \textit{David}, and \textit{Ricardo}, and they perform second best after models initialized at $p=1$ for \textit{Longdress}, \textit{Redandblack}, \textit{Loot}, and \textit{Soldier} with distortion metric YUV-PSNR. 

\subsection{Complexity Trade-off}
\begin{figure}[t]
    \centering
    \begin{subfigure}{0.4\textwidth}
        \includegraphics[width=\textwidth,trim=0.05in 0.0in 0.05in 0.0in,clip]{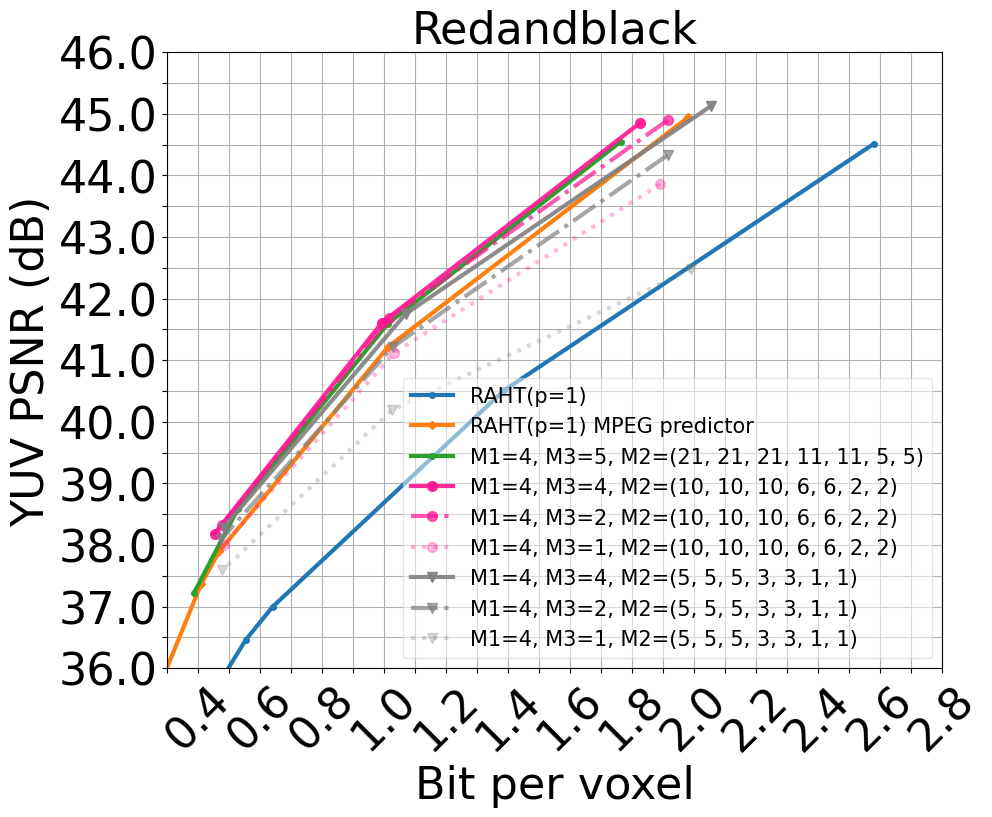}
    \end{subfigure}
    \caption{Complexity analysis curves \textit{Redandblack}}
    \vspace{-0.7cm}
    \label{fig:code_gain_performance_various_complex}
\end{figure}

In this subsection, we demonstrate RD performance under different computational complexity constraints, separately at the encoder and decoder, which we express as the number of network layers (unrolled iterations).
Recall that the computational complexity of the RD-optimized encoder is $O(M_1M_2M_3N)$ and that the computational complexity of the decoder is $O(M_1M_2N)$, where $M_1$ is the number of unrolled CGD iterations for the inverse of $\bPhi_l^\top\bPhi_l$, $M_2$ is the maximum degree of the Taylor expansion of the ortho-normalization operation at each level $l$, and $M_3$ is the number of unrolled PGD iterations in the encoder.
Fig.~\ref{fig:code_gain_performance_various_complex} compactly provides a wealth of information about the available tradeoffs.
While it does not explicitly investigate sensitivity of RD performance to $M_1$, we have found that $M_1=4$ CGD iterations are sufficient to robustly approximate the matrix inverse.  To avoid clutter we have set this to a constant in the figure.

First, the figure clearly shows that, at the decoder, more computation is better.
Specifically, it groups curves into low (gray) and high (pink) decoder complexity $O(M_1M_2N)$, where the high-complexity decoder has exactly twice the complexity of the low-complexity decoder.
Within each group, the difference in curves is the encoder complexity, with $M_3=1,2,4$.
Note that lower (resp.~higher) decoder complexity implies a synthesis transform that is further from (resp.~closer to) orthonormality.  Thus the gap between corresponding curves for low and high decoder complexity represents the cost due to non-orthonormality.  This cost can be over 1 dB at the higher bit rates, when the encoder complexity is low ($M_3=1$).

Second, the figure clearly shows that, at the encoder, more computation is better.
Specifically, within each group of curves for a fixed decoder complexity, the encoder complexity increases from $M_3=1$ to $M_3=4$.
It can be seen that the improvement is rapid from $M_3=1$ to $M_3=2$, but slows significantly from $M_3=2$ to $M_3=4$, whether the decoder complexity is low or high.  However, the improvements are smaller when the decoder complexity is high.

Third, the figure shows that \textit{complexity can be shifted between the encoder and decoder}.
In particular, for the low-complexity decoder (which is further from orthonormal), RD performance can be recovered almost fully by putting more computation into the encoder (increasing $M_3$); this is particularly clear in the low-bitrate region. Conversely, higher encoder complexity is not required if the decoder has higher complexity (closer to orthonormality).

\begin{figure}[t]
    \centering
    \begin{subfigure}{0.4\textwidth}
    \includegraphics[width=\textwidth,trim=0.05in 0.0in 0.05in 0.0in,clip]{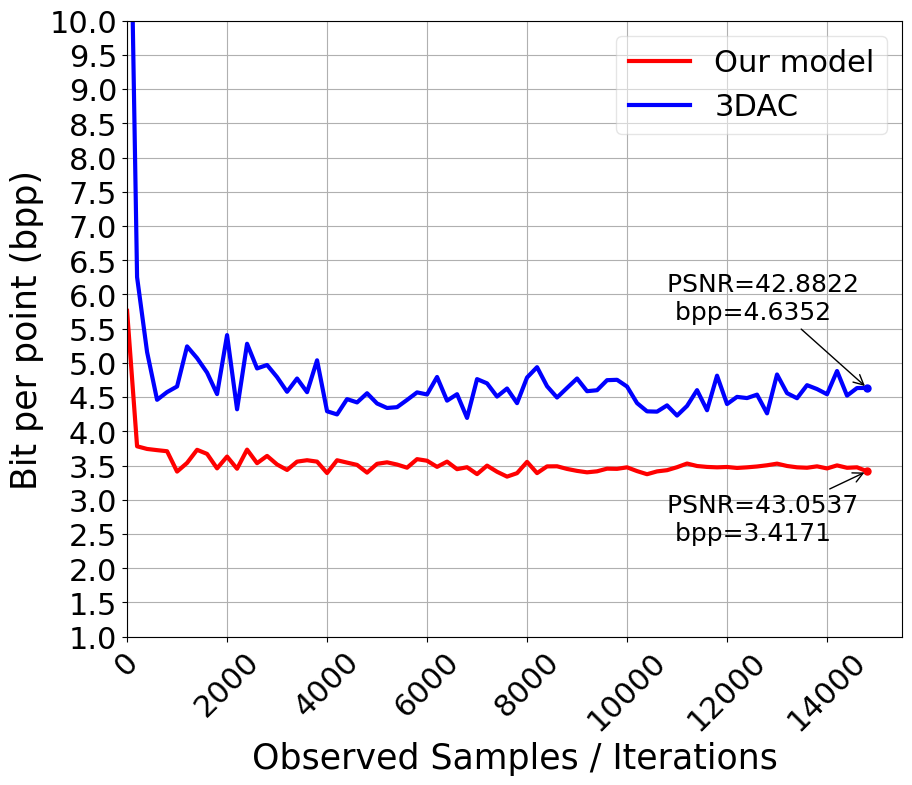}
    \end{subfigure}
    \vspace{-0.2cm}
    \caption{Evaluation of training checkpoints using \textit{Longdress}}
    \label{fig:overfit_scenerio}
\end{figure}

\subsection{Parameter comparisons}

Finally, we compare our model with others in terms of trainable parameter count.
Table~\ref{tab:parameter_comparison} shows the estimated number of parameters trained end-to-end on our system versus other recent systems \cite{fang20223dac,shao2024diffusion,mao2025spac,huo2025sptrans,mao2025pcacgan}.
Our parameters, listed fully in Tab.~\ref{tab:trainale_parameter}, are almost exclusively related to the decoder.  Our encoder has only a few parameters of its own, as it is essentially a function of the decoder --- specifically an unrolled optimization procedure for finding RD-optimal coefficients for the decoder.  The few learnable parameters in our encoder relate to controlling the convergence of the optimization procedure.
Our decoder, also, has orders of magnitude fewer coefficients than other approaches because it is a linear decoder based only on sparse convolution operators at each level of detail \textbf{without} cross-channels, multi-channels or non-linear activation functions, hence each convolutional operator in turn has only a small number of parameters. The few other learnable parameters relate to the tweakable parameters of the Taylor expansions of the inverse square root, the prediction module, and the entropy code.  All of our learned parameters initial conditions are set to intuitive defaults.  
This is the advantage of our approach, in which every element of our encoder and decoder is interpretable.

\begin{table}
\begin{center}

    \begin{tabular}{lll}
    \hline
        Scheme & Architecture & Params \\
        \hline
         Our paper &  Deep Unrolled RDO & $\approx 750$ \\
         3DAC (2022) \cite{fang20223dac} &  Deep Entropy Modeling & $\approx 10^5$ \\
         Diffusion-based (2024) \cite{shao2024diffusion} & Generative Diffusion & $\approx 10^6$ \\
         SPAC (2025) \cite{mao2025spac} & Hierarchical Autoencoder & $\approx 10^6$ \\
         SP-Trans (2025) \cite{huo2025sptrans} & Sparse Transformer & $\approx 10^7$ \\
         PCAC-GAN (2025) \cite{mao2025pcacgan} & Sparse Tensor GAN & $\approx 10^7$ \\
         \hline
    \end{tabular}
\caption{Comparison of Model Parameter Counts.}
\vspace{-0.7cm}
\label{tab:parameter_comparison}
\end{center}
\end{table}

With fewer parameters to train, and good default initial conditions for each, overfitting our system is difficult.  
Consequently, our system is robust to train/test distribution mismatch, and we are able to train our system to have good performance with less data than required by other approaches, as shown in Fig.\;\ref{fig:overfit_scenerio}.

\section{Conclusion}
\label{sec:conclude}
We unrolled the rate-distortion optimization algorithm into layers of a feed-forward network, resulting in an interpretable, variable complexity compression model that is also capable of adapting to data by training.
Experiments show that our model outperforms the MPEG G-PCC predictor by $6\%-11\%$ reduction in bit rate. 
While we focused narrowly on 3D point cloud attribute compression here, for future work, we will extend our deep unrolled RDO framework to more general 3D visual data types such as Gaussian Splat (representing radiance fields) using rendering-oriented distortion as briefly mentioned in Sec.\;\ref{sec:geometry_dependent_norm}.

\bibliographystyle{IEEEtran}
\bibliography{ref2}

\newpage
\clearpage
\section{Appendix}
\label{sec:appendix}
\subsection{Taylor expansion of $\mathbf{X}^{-1}$ and gradient descent}
\label{sec:taylor_inverse}

The Taylor expansion of $\mathbf{X}^{-1}$ to $t$ terms is equivalent to the solution of a minimum-norm problem by gradient descent in $t$ steps.
To see that, consider the following.
We can compute $\mathbf{X}^{-1}$ using a Taylor series expansion of $f(x)=x^{-1}$.  The Taylor series expansion of a function $f(x)$ around a point $x_0$ is
\begin{align}
    f(x) & =
    f(x_0) + f'(x_0)(x-x_0) + \frac{1}{2}f''(x_0)(x-x_0)^2  \nonumber \\
    & \mbox{} + \frac{1}{3!}f'''(x_0)(x-x_0)^3 + \cdots .
    \label{eqn:taylor_series}
\end{align}
In particular, for $f(x)=x^{-1}$, we have $f'(x)=-x^{-2}$, $f''(x)=2x^{-3}$, $f'''(x)=-3!x^{-4}$, etc.  Hence
\begin{eqnarray}
    \lefteqn{x^{-1}}
    & & = x_0^{-1} - x_0^{-2}(x-x_0) + x_0^{-3}(x-x_0)^2 + \cdots \nonumber \\
    & & = x_0^{-1} + x_0^{-2}(x_0-x) + x_0^{-3}(x_0-x)^2 + \cdots \nonumber \\
    & & = \frac{1}{x_0}\left[1 + \frac{x_0-x}{x_0} + \left(\frac{x_0-x}{x_0}\right)^2 + \cdots\right] \nonumber \\
    & & = \frac{1}{x_0}\left[1 + \left(1-\frac{x}{x_0}\right) + \left(1-\frac{x}{x_0}\right)^2 + \cdots\right] \nonumber \\
    & & = 2\mu\left[1 + \left(1-2\mu x\right) + \left(1-2\mu x\right)^2 + \cdots\right] ,
\end{eqnarray}
where $\mu=1/(2x_0)$.
The series converges if $x/x_0=2\mu x$ lies in $(0,2)$. 
Thus, if a symmetric positive definite matrix $\X$ has all its eigenvalues less than $1/\mu$, i.e., $\mu<1/\lambda_{\max}(\X)$, then $2\mu \X$ will have all its eigenvalues in $(0,2)$, and the Taylor series for $f(\mathbf{X})=\mathbf{X}^{-1}$ will converge. 
Thus, $\mathbf{X}^{-1}$ can be written as
\begin{equation}
    \mathbf{X}^{-1} = \lim_{t\rightarrow\infty} 2\mu \sum_{m=0}^t (I-2\mu \mathbf{X})^m
    \label{eqn:Xinv_taylor_series}
\end{equation}
as long as $\mu<1/\lambda_{\max}(\mathbf{X})$.

Alternatively, we can compute $\mathbf{X}^{-1}$ by minimizing over $\M$ the squared matrix norm $||\mathbf{X}\mathbf{M}-\I||_{P,Q}^2$, by gradient descent, where $\mathbf{X}$, $\P$, and $\Q$ are symmetric positive definite.
To see how to do this, first consider minimizing the squared $\ell_2$-norm $||\H \m-\b||_2^2$ for some vectors $\m$ and $\b$.  Since
\begin{align}
    ||\H \m-\b||_2^2
    & = (\m^\top \H^\top-\b^\top)(\H \m-\b) \\
    & = \m^\top \H^\top \H \m-2\m^\top \H^\top \b+\b^\top \b ,
\end{align}
the gradient of $||\H \m-\b||_2^2$ with respect to $\m$ is
\begin{equation}
    \nabla_\m||\H \m-\b||_2^2 = 2\H^\top \H \m - 2\H^\top \b .
\end{equation}
Thus, to find the vector $\m$ that minimizes $||\H \m-\b||_2^2$, we can perform gradient descent from an initial guess $\m^{(0)}=0$ using stepsize $\mu$ as
\begin{align}
    \m^{(t+1)}
    & = \m^{(t)} - \mu (2\H^\top \H \m^{(t)} - 2\H^\top \b) \\
    & = \m^{(t)} - 2\mu \H^\top(\H \m^{(t)} - \b) \\
    & = 2\mu \H^\top \b + (I - 2\mu \H^\top \H)\m^{(t)} \\
    & = 2\mu \sum_{p=0}^t (I - 2\mu \H^\top \H)^p \H^\top \b .
\end{align}
This will converge as long as the eigenvalues of $\H^\top \H$ lie in $(0,1/\mu)$, in other words, as long as $\mu<1/\lambda_{\max}(\H^\top \H)$.

We can use this to operate on many vectors at the same time.  For square matrices $\B$ and $\M$, we can minimize the squared Frobenius norm $||\H\M-\B||_F^2$ using gradient descent:
\begin{align}
    \M^{(t+1)}
    & = \M^{(t)} - \mu (2\H^\top \H \M^{(t)} - 2\H^\top \B) \\
    & = \M^{(t)} - 2\mu \H^\top(\H \M^{(t)} - \B) \\
    & = 2\mu \H^\top \B + (\I - 2\mu \H^\top \H)\M^{(t)} \\
    & = 2\mu \sum_{p=0}^t (\I - 2\mu \H^\top \H)^p \H^\top \B .
    \label{eqn:Xinv_gradient_descent0}
\end{align}
This will converged as long as the eigenvalues of $\H^\top \H$ lie in $(0,1/\mu)$, or $\mu<1/\lambda_{\max}(\H^\top \H)$.

Now, to minimize
\begin{align}
    ||\X \M-\I||_{\P,\Q}^2=||\Q^{1/2}(\X\M-\I)\P^{-1/2}||_F^2 ,    
\end{align}
we choose $\P=\I$ and $\Q=\X^{-1}$.  Then
\begin{align}
    ||\X\M-\I||_{\P,\Q}^2=||\X^{1/2}\M-\X^{-1/2}||_F^2 . 
\end{align}
Thus (\ref{eqn:Xinv_gradient_descent0}) reduces to (\ref{eqn:Xinv_taylor_series}) if $\H=\X^{1/2}$ and $\B=\X^{-1/2}$.
The approximation converges as long as the eigenvalues of $\X=\H^T\H$ lie in $(0,1/\mu)$, i.e., $\mu<1/\lambda_{\max}(\X)$.

\subsection{Unrolling conjugate gradient descent}
\label{sec:unroll_cgd}

The linear systems that we need to solve --- \eqref{eqn:lowpass_linear_system} and \eqref{eq:Z_operation} --- have the form
\begin{align}
\boldsymbol{\X}F = \Tilde{F} .
\label{eq:linear_system}
\end{align}
This can be solved by minimizing the squared $\Q$-norm
\begin{align}
    ||\X F-\tilde F||_\Q^2
    & = (F^\top\X^\top-\tilde F^\top)\Q(\X F-\tilde F) \label{eqn:squared_Q_norm} \\
    & = F^\top\X^\top\Q\X F -2\tilde F^\top\Q\X F + \tilde F^\top\Q\tilde F \nonumber
\end{align}
for any symmetric positive definite matrix $\Q$.
Given that $\X=(\bPhi_l^\top\bPhi_l)$ is symmetric positive definite, we may solve \eqref{eq:linear_system} by minimizing \eqref{eqn:squared_Q_norm} with $\Q=\X^{-1}$, or equivalently minimizing the quadratic form
\begin{equation}
    Q(F) = \frac{1}{2} F^\top \X F - \Tilde{F}^\top F ,
\end{equation}
which has gradient
\begin{equation}
\nabla_FQ(F) = \X F - \Tilde{F} .
\end{equation}
Simple gradient descent has the following update rules, where $\alpha_t$ is the stepsize or {\em learning rate}:
\begin{align}
    \g^t &= \X F^{t} - \Tilde{F} = \g^{t-1} - \alpha_{t-1}\X \g^{t-1} \\
    F^{t+1} &= F^{t} - \alpha_t \g^t .
\end{align}
Adding a momentum factor $\beta_t$ and cumulative gradients $\v^t$ results in the well-known Accelerated Gradient Descent \red{\cite{nesterov2013introductory}}, also known as Conjugate Gradient Descent, with update rules
\begin{align}
    F_t & = F_{t-1} - \alpha_{t-1} \v_{t-1} \\
    \g_t & = \g_{t-1} - \alpha_{t-1} \X \v_{t-1} \\
    \v_t & = \g_t + \beta_{t-1} \v_{t-1} .
\end{align}
where both $\alpha_t$ and $\beta_t$ in each iteration $t$ may be trainable parameters.


\subsection{Taylor expansion of $\X^{-1/2}$}
\label{sec:taylor_sqrt_inverse}

For $f(x)=x^{-1/2}$, we have $f'(x)=(-1/2)x^{-3/2}$, $f''(x)=(3/4)x^{-5/2}$, $f'''(x)=(-15/8)x^{-7/2}$, etc.  Hence the Taylor series of $x^{-1/2}$ around $a$ is given in Fig.~\ref{fig:taylor_sqrt_inverse_equation}, where $\mu=1/(2a)$.


\begin{figure*}[!h]
\hrulefill
\begin{align}
    x^{-\frac{1}{2}} &= a^{-\frac{1}{2}} + \frac{(-\frac{1}{2})a^{-\frac{3}{2}}}{1!}(x - a) 
    + \frac{(-\frac{1}{2})(-\frac{3}{2})a^{-\frac{5}{2}}}{2!}(x - a)^2
    + \frac{(-\frac{1}{2})(-\frac{3}{2})(-\frac{5}{2})a^{-\frac{7}{2}}}{3!}(x - a)^3
    + ... \\
    &= a^{-\frac{1}{2}} \left [ 1 + \frac{(\frac{1}{2})a^{-1}}{1!}(a - x) 
    +\frac{(\frac{1}{2})(\frac{3}{2})a^{-2}}{2!}(a - x)^2 
    +\frac{(\frac{1}{2})(\frac{3}{2})(\frac{5}{2})a^{-3}}{3!}(a - x)^3 + ...\right ] \\
    &= a^{-\frac{1}{2}} \left [ 1 + \frac{(\frac{1}{2})}{1!}\left (1 - \frac{x}{a} \right )
    + \frac{(\frac{1}{2})(\frac{3}{2})}{2!}\left (1 - \frac{x}{a} \right )^2  
    + \frac{(\frac{1}{2})(\frac{3}{2})(\frac{5}{2})}{3!}\left (1 - \frac{x}{a} \right )^3 + ...\right ] \\
    &= (\frac{1}{a})^{\frac{1}{2}} \left [ 1 + \frac{1}{2 * 1!}\left (1 - \frac{x}{a} \right )
    + \frac{1 * 3}{2^2 * 2!}\left (1 - \frac{x}{a} \right )^2  
    + \frac{1*3*5}{2^3*3!}\left (1 - \frac{x}{a} \right )^3 + ...\right ]  \\
    &= \sqrt{2\mu} \left [ 1 + \frac{1}{2 * 1!}\left (1 - 2\mu x \right )
    + \frac{1 * 3}{2^2 * 2!}\left (1 - 2\mu x \right )^2  
    + \frac{1*3*5}{2^3*3!}\left (1 - 2\mu x \right )^3 + ...\right ] 
\end{align}
\hrulefill
\vspace*{4pt}
\caption{Taylor series for $x^{-1/2}$ around $x=a=1/(2\mu)$.}
\label{fig:taylor_sqrt_inverse_equation}
\end{figure*}

The series converges if $x/a=2\mu x$ lies in $(0,2)$.  Thus, if a symmetric positive definite matrix $\X$ has all its eigenvalues less than $1/\mu$, i.e., $\mu<1/\lambda_{\max}(\X)$, then $2\mu \X$ will have all its eigenvalues in $(0,2)$, and the Taylor series for $f(\X)=\X^{-1/2}$ will converge.  Thus, $\X^{-1/2}$ can be written as
\begin{equation}
   \X^{-1/2} = \sqrt{2\mu} \left [ 1 + \sum_{n=1}^\infty \frac{1...(2n-1)}{2^n n!}\left (1 - 2\mu \X \right )^n   \right] 
\end{equation}
as long as $\mu<1/\lambda_{\max}(\X)$.

\subsection{Expression for $\mathbf{\Psi}_l^\top\mathbf{\Psi}_l$}
\label{sec:Psi_Psi}

\begin{eqnarray}
\mathbf{X}
& = & \mathbf{\Psi}_l^T\mathbf{\Psi}_l \\
& = & 
\mathbf{Z}_l\mathbf{\Phi}_{l+1}^T\mathbf{\Phi}_{l+1}\mathbf{Z}_l^T \\
& = & \mathbf{I}_l^b\left[\mathbf{I}_{l+1}-(\mathbf{\Phi}_{l+1}^T\mathbf{\Phi}_{l+1})\mathbf{A}_l^T(\mathbf{\Phi}_l^T\mathbf{\Phi}_l)^{-1}\mathbf{A}_l\right] \nonumber \\
& & \times(\mathbf{\Phi}_{l+1}^T\mathbf{\Phi}_{l+1}) \nonumber \\
& & \times\left[\mathbf{I}_{l+1}-\mathbf{A}_l^T(\mathbf{\Phi}_l^T\mathbf{\Phi}_l)^{-1}\mathbf{A}_l(\mathbf{\Phi}_{l+1}^T\mathbf{\Phi}_{l+1})\right]\mathbf{I}_l^{bT} \\
& = & \mathbf{I}_l^b\left[(\mathbf{\Phi}_{l+1}^T\mathbf{\Phi}_{l+1})\right. \nonumber \\
& & -2(\mathbf{\Phi}_{l+1}^T\mathbf{\Phi}_{l+1})\mathbf{A}_l^T(\mathbf{\Phi}_l^T\mathbf{\Phi}_l)^{-1}\mathbf{A}_l(\mathbf{\Phi}_{l+1}^T\mathbf{\Phi}_{l+1}) \nonumber \\
& & +(\mathbf{\Phi}_{l+1}^T\mathbf{\Phi}_{l+1})\mathbf{A}_l^T(\mathbf{\Phi}_l^T\mathbf{\Phi}_l)^{-1}\mathbf{A}_l(\mathbf{\Phi}_{l+1}^T\mathbf{\Phi}_{l+1}) \nonumber \\
& & \;\;\;\left.\mathbf{A}_l^T(\mathbf{\Phi}_l^T\mathbf{\Phi}_l)^{-1}\mathbf{A}_l(\mathbf{\Phi}_{l+1}^T\mathbf{\Phi}_{l+1})\right]\mathbf{I}_l^{bT} \\
& = & \mathbf{I}_l^b\left[(\mathbf{\Phi}_{l+1}^T\mathbf{\Phi}_{l+1})\right. \nonumber \\
& & -2(\mathbf{\Phi}_{l+1}^T\mathbf{\Phi}_{l+1})\mathbf{A}_l^T(\mathbf{\Phi}_l^T\mathbf{\Phi}_l)^{-1}\mathbf{A}_l(\mathbf{\Phi}_{l+1}^T\mathbf{\Phi}_{l+1}) \nonumber \\
& & +(\mathbf{\Phi}_{l+1}^T\mathbf{\Phi}_{l+1})\mathbf{A}_l^T(\mathbf{\Phi}_l^T\mathbf{\Phi}_l)^{-1}(\mathbf{\Phi}_l^T\mathbf{\Phi}_l) \nonumber \\
& & \;\;\;\left.(\mathbf{\Phi}_l^T\mathbf{\Phi}_l)^{-1}\mathbf{A}_l(\mathbf{\Phi}_{l+1}^T\mathbf{\Phi}_{l+1})\right]\mathbf{I}_l^{bT} \\
& = & \mathbf{I}_l^b\left[(\mathbf{\Phi}_{l+1}^T\mathbf{\Phi}_{l+1})\right. \nonumber \\
& & \left.-(\mathbf{\Phi}_{l+1}^T\mathbf{\Phi}_{l+1})\mathbf{A}_l^T(\mathbf{\Phi}_l^T\mathbf{\Phi}_l)^{-1}\mathbf{A}_l(\mathbf{\Phi}_{l+1}^T\mathbf{\Phi}_{l+1})\right]\mathbf{I}_l^{bT} .
\end{eqnarray}

\subsection{Rate term}
\label{sec:rate_term}


The rate terms come from assuming that the coefficients are zero-mean Laplacian random variables $X$ with mean absolute value $b=E|X|$, i.e., $p(x)=(1/2b)\exp\{-|x|/b\}$, so that for $t>0$, $P\{X\geq t\}=(1/2)\exp\{-t/b\}$, and for $|t|\geq\Delta/2$,
\begin{eqnarray}
    \lefteqn{P\{X\in[t-\Delta/2,t+\Delta/2]\}} \\
    & = & P\{X\geq|t|-\Delta/2\} - P\{X\geq|t|+\Delta/2\} \\
    & = & \frac{1}{2}\left(e^{-\frac{|t|-\Delta/2}{b}} - e^{-\frac{|t|+\Delta/2}{b}}\right) \\
    & = & \frac{1}{2}e^{-\frac{|t|}{b}}\left(e^{\frac{\Delta}{2b}} - e^{-\frac{\Delta}{2b}}\right) \\
    & \approx & \frac{1}{2}e^{-\frac{|t|}{b}}\left((1+\frac{\Delta}{2b})-(1-\frac{\Delta}{2b})\right) \\
    & = & \frac{\Delta}{2b}e^{-\frac{|t|}{b}} ,
\end{eqnarray}
whence the number of bits to encode the event $\{X\in[t-\Delta/2,t+\Delta/2]\}$ is approximately
\begin{equation}
    -\log_2\left(\frac{\Delta}{2b}e^{-\frac{|t|}{b}}\right)
    = \frac{|t|}{b\ln 2} - \log_2\frac{\Delta}{2b} .
\end{equation}
(The rate term should never be negative, hence this expression is sometimes clipped to remain above zero.)  For vectors of such Laplacians, such as $G_l$, the total rate to encode $\hat G_l$ is
\begin{equation}
    rate(\hat G_l)
    = \left(\frac{||G_l||_1}{b_l\ln 2} - \log_2\frac{\Delta}{2b_l}\right) ,
\end{equation}
where $b_l$ is the scale parameter for the Laplacian.

\subsection{Change of basis}
\label{sec:basis_changes}
We first give an example of the transformation $\M^{(3)}$ for a simple case of 3 levels, $\cF_1 = \cF_{0} \oplus \cG_{0} \oplus \cG_{1}$

\begin{align}
    \begin{bmatrix}
        \overline{G}_{1} \\
        \overline{G}_{0} \\
        \overline{F}_{0}
    \end{bmatrix} = \M^{(3)} \begin{bmatrix}
        \overline{G}^{''}_{1} \\
        \overline{G}^{''}_{0} \\
        \overline{F}_{0}
    \end{bmatrix} 
\end{align}
where the last 2 rows of $\M^{(3)}$ is already established in Section~\ref{sec:prediction},
\begin{align}
    \M^{(3)} = \begin{bmatrix}
        \I &\M^{(3)}_{0, 1} &\M^{(3)}_{0, 2} \\
        \mathbf{0} &\I &\R_{\bPsi_{0}} \mathbf{Z}_{0} (\bPhi_{1}^\top\bPhi_{1}) (\P^\top_{0} - \A^\top_{0})\R_{\bPhi_{0}} \\
        \mathbf{0} &\mathbf{0} &\I
    \end{bmatrix}
\end{align}
the first row can be filled by finding $\M_{0, 1}$ and $\M_{0, 2}$, which are required to satisfy,
\begin{align}
    \overline{G}^{'}_{1} = \begin{bmatrix}
        \M^{(3)}_{0, 1} &\M^{(3)}_{0, 2}
    \end{bmatrix} \begin{bmatrix}
        \overline{G}^{''}_{0} \\
        \overline{F}_{0}
    \end{bmatrix}  .
\end{align}
It is clear that calculation of $\overline{G}^{'}_{1}$ required a sequence of operations, and $\M_{0, 1}$ and $\M_{0, 2}$ are the results of this sequence of operations:
\begin{align}
    &\begin{bmatrix}
        \M^{(3)}_{0, 1} &\M^{(3)}_{0, 2}
    \end{bmatrix} = \A \times \B  \times \C \\
    &\A = \R_{\bPsi_{1}} \mathbf{Z}_{1} (\bPhi_{2}^\top\bPhi_{2})(\P^\top_{1} - \A^\top_{1}) \\
    &\B = \begin{bmatrix}
         \mathbf{Z}_{0}^\top \R_{\bPsi_{0}} &\A^\top_{0}\R_{\bPhi_{0}}
    \end{bmatrix} \\ 
    &\C = \begin{bmatrix}
        \I &\R_{\bPsi_{0}} \mathbf{Z}_{0} (\bPhi_{1}^\top\bPhi_{1})(\P^\top_{0} - \A^\top_{0})\R_{\bPhi_{0}} \\
        \mathbf{0} &\I
    \end{bmatrix}
\end{align}
where it is obvious that operation $\C$ is itself the transformation for 2 levels case, so that $\C = \M^{(2)}$, and 
\begin{align}
    \overline{G}^{'}_{1} &= \A F_{1} = \A \times \B \begin{bmatrix}
        \overline{G}_{0} \\
        \overline{F}_{0}
    \end{bmatrix}
\end{align}

We have established the rules to further derive the change of basis transformation for more levels.
Let the next row be defined as follows
\begin{align}
    \overline{G}^{'}_{l} = 
    \A^{(l-1)} \times \B^{(l-1)}  \times \M^{(l-1)}
    \begin{bmatrix}
        \overline{G}^{''}_{l-1}\\
        \cdots \\
        \overline{G}^{''}_{0} \\
        \overline{F}_{0}
    \end{bmatrix} 
\end{align}
where $\M^{(l-1)}$ and $\B^{(l-1)}$ satisfy 
\begin{align}
    F_{l} &= \B^{(l-1)} \begin{bmatrix}
        \overline{G}_{l-1}\\
        \cdots \\
        \overline{G}_{0} \\
        \overline{F}_{0}
    \end{bmatrix} \\ 
    \begin{bmatrix}
        \overline{G}_{l-1}\\
        \cdots \\
        \overline{G}_{0} \\
        \overline{F}_{0}
    \end{bmatrix}  &= \M^{(l-1)} \begin{bmatrix}
        \overline{G}^{''}_{l-1}\\
        \cdots \\
        \overline{G}^{''}_{0} \\
        \overline{F}_{0}
    \end{bmatrix} 
\end{align}
and 
\begin{align}
    \A^{(l-1)} &= \R_{\bPsi_{l}} \mathbf{Z}_{l} (\bPhi_{l+1}^\top\bPhi_{l+1})(\P^\top_{l} - \A^\top_{l})
\end{align}
Then, 
\begin{align}
    \M^{(l)} = \left [ \begin{array}{cc}
    \I &\A^{(l-1)} \times \B^{(l-1)}  \times \M^{(l-1)} \\
    \mathbf{0} &\M^{(l-1)}
    \end{array}\right]
\end{align}

\subsection{Using $\M^{-1}$ instead of $\M^\top$ in PGD}
\label{sec:PGD_using_M_inverse}

We seek to find the real vector
\begin{align}
    V_P^*
    & = \argmin_{V_P} \left[||f-\bTheta \M V_P||^2 + \lambda||\bGamma_P V_P||_1 \right].
    \label{eqn:encoder_objective_relaxed2}
\end{align}
We let $V=\M V_P$ and define $h(V)=||f-\bTheta V||^2$.  Note that for sufficiently small $\mu$, and any $V^{(t)}$, $h(V)$ is upper-bounded (majorized) at all $V$ by
\begin{align}
    h_\mu(V,V^{(t)}) \stackrel{\Delta}{=} & \; h(V^{(t)}) + \nabla h(V^{(t)})^\top(V-V^{(t)}) \nonumber \\
    & +\frac{1}{2\mu}||V-V^{(t)}||_2^2 ,
    \label{eqn:upper_bound}
\end{align}
if $\mu$ is sufficiently small.
In turn,
\begin{align}
    & \mu h_\mu(V,V^{(t)}) \nonumber \\
    & = \mu\nabla h(V^{(t)})^{\top} V + \frac{1}{2}||V||^2 - V^{(t)\top} V + K' \\
    & = \frac{1}{2}||V - (V^{(t)}-\mu\nabla h(V^{(t)})||_2^2 + K'' \\
    & \leq \frac{c}{2}||\M^{-1}\left[V - (V^{(t)}-\mu\nabla h(V^{(t)})\right]||_2^2 + K'' 
\end{align}
where $K'$ and $K''$ depend on $V^{(t)}$ but not on $V$, and $c=||\M||^2$.
Thus, \eqref{eqn:encoder_objective_relaxed2} may be solved by the minimization-maximization algorithm,
\begin{align}
    V_P^{(t+1)} = \argmin_{V_P} \left[\frac{1}{2}||V_P - U_P^{(t)}||^2 + \frac{\mu}{c}\lambda||\bGamma_P V_P||_1 \right] ,
\end{align}
where 
\begin{align}
    U_P^{(t)} &= \M^{-1} (V^{(t)}- \mu \nabla h(V^{(t)})) \\
    &= V_P^{(t)} - \mu \M^{-1} \nabla h(\M V_P^{(t)}) \\
    &= V_P^{(t)} - \mu \M^{-1} \bTheta^\top(f-\bTheta \M V_P^{(t)} )
\end{align}

\end{document}